\documentclass[onefignum,onetabnum]{siamart190516}

\usepackage{amsfonts}
\usepackage{graphicx,subfigure}
\usepackage{epstopdf}
\usepackage{algorithmic}
\usepackage{amsopn}
\newsiamremark{remark}{Remark}
\newsiamremark{hypothesis}{Hypothesis}
\newsiamthm{claim}{Claim}

\headers{Stochastic Peierls-Nabarro Model for Dislocations in HEAs}{Tianpeng Jiang, Yang Xiang, and Luchan Zhang}

\title{Stochastic Peierls-Nabarro Model for Dislocations in High Entropy Alloys}

\author{Tianpeng Jiang\thanks{Department of Mathematics, The Hong Kong University of Science and Technology, Clear Water Bay, Kowloon, Hong Kong SAR (\email{tjiangad@connect.ust.hk}, \email{maxiang@ust.hk}, \email{malczhang@ust.hk}).}
	\and Yang Xiang\footnotemark[1] \and Luchan Zhang\footnotemark[1]}
\begin{document}

\maketitle

% REQUIRED
\begin{abstract}
High entropy alloys (HEAs) are single phase crystals that consist of random solid solutions of multiple elements  in approximately equal proportions. This class of novel materials have exhibited superb mechanical properties, such as high strength combined with other desired features. The strength of crystalline materials is associated with the motion of dislocations. In this paper, we  derive a stochastic continuum model based on the Peierls-Nabarro framework for inter-layer dislocations in a bilayer
HEA from an atomistic model that incorporates the atomic level randomness. We use asymptotic analysis and  limit theorem in the convergence from the atomistic model to the continuum model.  The total energy in the  continuum model consists of a stochastic elastic energy in the two layers,  and a stochastic misfit energy that accounts for the inter-layer nonlinear interaction.  The obtained continuum model can be considered as a stochastic generalization of the classical, deterministic Peierls-Nabarro model for the dislocation core and related properties.
This derivation also validates the stochastic model adopted by Zhang $et$ $al.$ (Acta Mater. 166, 424-434, 2019).
\end{abstract}

% REQUIRED

% REQUIRED

%74A25   Molecular, statistical, and kinetic theories
%74A40   Random materials and composite materials
%74B99 Mechanics of deformable solids None of the above, but in this section
%49K45  Calculus of variations and optimal control; optimization  Problems involving randomness [See also 93E20]
%35R60   Partial differential equations with randomness, stochastic partial differential equations [See also 60H15]
%60H15   Stochastic partial differential equations

\section{Introduction}
Different from the conventional alloys developed based on one primal element, high entropy alloys (HEAs) are single phase crystals that consist of random solid solutions of multiple elements (five or more) in approximately equal proportions \cite{HEAs2004_1,HEAs2004_2,reviewHEAs2014,reviewHEAs2014_2,fracture_resis_2,reviewHEAs2017,curtin_review2020}. Because each lattice site in HEAs is randomly occupied by one of the main elements,  HEAs have significantly higher mixing entropies than those in conventional alloys. It is widely believed that the high mixing entropies in these materials facilitate the formation of  simple structures (e.g., face-centered cubic or body-centered cubic lattices) and enable many ideal engineering properties, such as  high temperature stability, high strength, high fracture resistance, and high radiation-damage resistance, etc. Because of these promising properties, HEAs have attracted considerable research interest ever since the discovery of this novel class of materials.
One attractive mechanical property of HEAs is the high strength combined with high ductility and other desired features, which cannot be achieved in single-component crystals and conventional alloys. There are extensive experimental studies (e.g., \cite{senkov2011,otto2013,Yoshida2019}) and atomistic simulations/{\it ab initio} studies (e.g., \cite{tamm2015,sharma2016,kormann2017,RaoCurtin2017,wangyz2018,Farkas2020}) available on the high strength of HEAs (see also the reviews \cite{reviewHEAs2014,reviewHEAs2014_2,fracture_resis_2,reviewHEAs2017,curtin_review2020}).

Theoretically,
the strength of crystalline materials is determined by the motion of dislocations (line defects) \cite{dislocation_book}.
Many of the existing models for the strength of HEAs are based on the classical ideas of solute solution strengthening; e.g., the Labusch model~\cite{labusch1970statistical}.
While the original Labusch model is directly applicable for cases where there is a distinction between solute and solvent atoms in conventional alloys (unlike in HEAs), some extensions to the HEA case have focused on how to combine contributions from each component to the strength.
Toda-Caraballo {\it et al.}~\cite{toda2015modelling} adopted an averaging procedure for this purpose.
Curtin {\it et al.}~\cite{varvenne2016theory,varvenne2017solute,Curtin2020} explicitly considered the interaction energy between a solute atom and a dislocation in a matrix that was described as an effective medium with random local concentration fluctuations.

Recently,  Zhang \textit{et al.} \cite{theoretical1}  have developed a stochastic continuum model under the framework of the Peierls-Nabarro  model \cite{PN1,PN2,dislocation_book}  to understand how random site occupancy
affects intrinsic strength of HEA materials. The stochastic Peierls-Nabarro model accounts for the randomness  and short-range order on the atomic level in HEAs. Nonlinear effect associated with the dislocation core is described by a stochastic nonlinear interplanar potential.
 The model predicts the intrinsic strength of
HEAs as a function of the standard deviation and the correlation length of the randomness. They also found that the compositional randomness in an HEA gives significant rise to the intrinsic
strength, which agrees with atomistic simulations and experiments.

Despite the success of these theories in predicting results that agree with those of experiments and atomistic simulations, convergence from atomistic models to these theories has not been examined in the literature. The theory in Ref.~\cite{toda2015modelling} focuses on averaging the result of the Labusch model and does not explicitly consider the elastic interaction of dislocations with the atomic level randomness in HEAs. The theories in Ref.~\cite{varvenne2016theory,varvenne2017solute,Curtin2020} were derived from continuum models of interactions under linear elasticity theory; as a result, these models may not necessarily  accurately incorporate the influence of the atomic level randomness on the dislocation core, in which linear elasticity theory does not apply. In Ref.~\cite{theoretical1}, the stochastic effects in the nonlinear interaction under the Peierls-Nabarro model are incorporated phenomenologically instead of direct derivation from the atomistic model.

In this paper, we derive a continuum model for inter-layer dislocations in a bilayer HEA from an atomistic model that incorporates the atomic level randomness. The continuum model is under the framework of the Peierls-Nabarro model, in which the nonlinear effect within the dislocation core region is included.
The total energy in the obtained stochastic continuum model consists of a stochastic elastic energy in the two layers,  and a stochastic misfit energy that accounts for the nonlinear inter-layer interaction  and whose energy density is the stochastic generalized stacking fault energy (or the $\gamma$-surface).  The obtained continuum model can be considered as a stochastic generalization of the classical, deterministic Peierls-Nabarro model \cite{PN1,PN2,dislocation_book} with generalized stacking fault energy \cite{vitek}.
This derivation also validates the stochastic model adopted in Ref.~\cite{theoretical1}.

We use asymptotic analysis and  (modified) central limit theorem in the convergence from the atomistic model to the continuum model. The atomic level randomness is incorporated by assuming that each lattice site is occupied by atom species with certain distributions. In the derivation, we introduce a supercell whose size is much greater than the lattice constant, and in the meantime, much smaller than the length unit of the continuum model, and employ the Cauchy-Born rule \cite{BornHuang1954} for the derivation of the continuum formulation of the elastic energy and definition of the generalized stacking fault energy  \cite{vitek} for the calculation of the misfit energy.

The rest of the paper is organized as follows.  In Sec.~\ref{sec:PN}, we review the classical Peierls-Nabarro model for dislocations. In Sec.~\ref{section_setting}, we introduce the atomistic model for a bilayer HEA, from which the continuum model will be derived.  In Sec.~\ref{section_misfit},
 we first calculate the generalized stacking fault energy of the bilayer HEA using the atomistic model, and then derive stochastic continuum formulations for it and the misfit energy.
  In Sec.~\ref{section_elstic}, we derive stochastic continuum formulation of the energy due to the intra-layer elastic interaction of the bilayer HEA from the atomistic model. In Sec.~\ref{section_total},
    we formulate the continuum stochastic total energy that incorporates the covariance of the randomness in the misfit and elastic energies, and rigorously prove the convergence from the atomistic model by modified central limit theorem. The stochastic model adopted in Ref.~\cite{theoretical1} is examined.
   In  Sec.~\ref{section_summary}, we summarize the results.

\section{Review of classical Peierls-Nabarro model} \label{sec:PN}

The Peierls-Nabarro  model for dislocations \cite{PN1,PN2,vitek,dislocation_book} is a continuum model that combines a long-range elastic field of a dislocation and an atomic-level description of its core.
In its classical form, it describes a straight dislocation with its core spread over a small, finite region along the slip plane.

We assume that there is an edge dislocation located along the $z$-axis, its Burgers vector $\mathbf b$ is in the $+x$-axis, and the $y=0$ plane is the slip plane.
The slip plane separates two linear elastic continua ($y>0$ and $y<0$). Across the slip plane $y=0$, there is a jump in the displacement in the $x$ direction, which is called  disregistry across the slip plane (i.e., slip in the $x$-direction). The disregistry function $\phi(x)=u^+(x)-u^-(x)$, where $u^+(x)$ and $u^-(x)$ are respectively the displacements in the $x$ direction on the atomic layers right above and below the slip plane,
and $\phi(-\infty)=0$, $\phi(+\infty)=b$, where $b$ is the length of the Burgers vector $\mathbf b$.
The Burgers vector distribution is $\rho(x)=\phi'(x)$, which characterizes the dislocation core and takes the form of a regularized delta-function. See Fig.~\ref{latticeDislocation}(b) for a schematic illustration of the disregistry function $\phi(x)$.

The total energy of a dislocation in the Peierls-Nabarro model can be written as
\begin{equation}\label{eqn:Etotal}
E_{\text{total}}=E_{\text{elastic}}+E_{\text{misfit}},
\end{equation}
where $E_{\text{elastic}}$ is the elastic energy in the upper and lower continua delimited by the slip plane and $E_{\text{misfit}}$ is the misfit energy associated with the nonlinear atomic interactions across the slip plane.

The misfit energy can be written in terms of the disregistry:
\begin{equation}\label{eqn:Emisfit}
E_{\text{misfit}}=\int_{-\infty}^{+\infty} \gamma(\phi(x))dx,
\end{equation}
where $\gamma(\phi)$ is the nonlinear interplanar potential.
In the classical Peierls-Nabarro model, $\gamma(\phi)$ is approximated by the Frenkel sinusoidal potential~\cite{Frenkel1926,dislocation_book},
\begin{equation}\label{eqn:Frenkel}
\gamma(\phi)=\frac{\mu b^2}{4\pi^2d}\left(1-\cos \frac{2\pi\phi}{b}\right),
\end{equation}
where $\mu$ is the shear modulus, and $d$ is the atomic interplanar spacing perpendicular to the slip plane. In general, the nonlinear potential $\gamma(\phi)$ is the generalized stacking fault energy (or the $\gamma$-surface) \cite{vitek} that is defined as the energy increment per unit length
when there is a uniform shift of $\phi$ between the upper and lower halves of a perfect lattice along the slip plane. See Sec.~\ref{subsec:gammasurface} and Fig.~\ref{latticeDislocation}(d) for more details of the generalized stacking fault energy   of a bilayer system.

In the case of a bilayer system with an inter-layer edge dislocation being considered in this paper, the elastic energy due to the intra-layer elastic interaction is
\begin{equation}\label{eqn:Eelastic}
E_{\text{elastic}}=\int_{-\infty}^{+\infty} \left[\frac{1}{2}\alpha \left(\frac{d u^+(x)}{dx}\right)^2+ \frac{1}{2}\alpha \left(\frac{d u^-(x)}{dx}\right)^2 \right]dx,
\end{equation}
where $\alpha$ is an elastic constant. Note that an edge dislocation in a three-dimensional space is considered in the classical Peierls-Nabarro model  \cite{PN1,PN2,dislocation_book}, with
the elastic energy
$E_{\text{elastic}}=\frac{1}{2}\int_{-\infty}^{+\infty}\sigma_{xy}(x)\phi(x)dx$,
where the shear stress on the slip plane is
$\sigma_{xy}(x)=\frac{\mu}{2\pi(1-\nu)}\int_{-\infty}^{+\infty}\frac{\phi'(x_1)}{x-x_1}dx_1$ ($\nu$ is the Poisson ratio). For a bilayer system, when the Frenkel sinusoidal potential in Eq.~\eqref{eqn:Frenkel} is used for the misfit energy, together with the elastic energy in Eq.~\eqref{eqn:Eelastic}, the model is the  Frenkel-Contorova model \cite{FKmodel}.

\section{Atomistic model of HEAs}\label{section_setting}

\begin{figure}[htbp]
\centering
    \includegraphics[width=\linewidth]{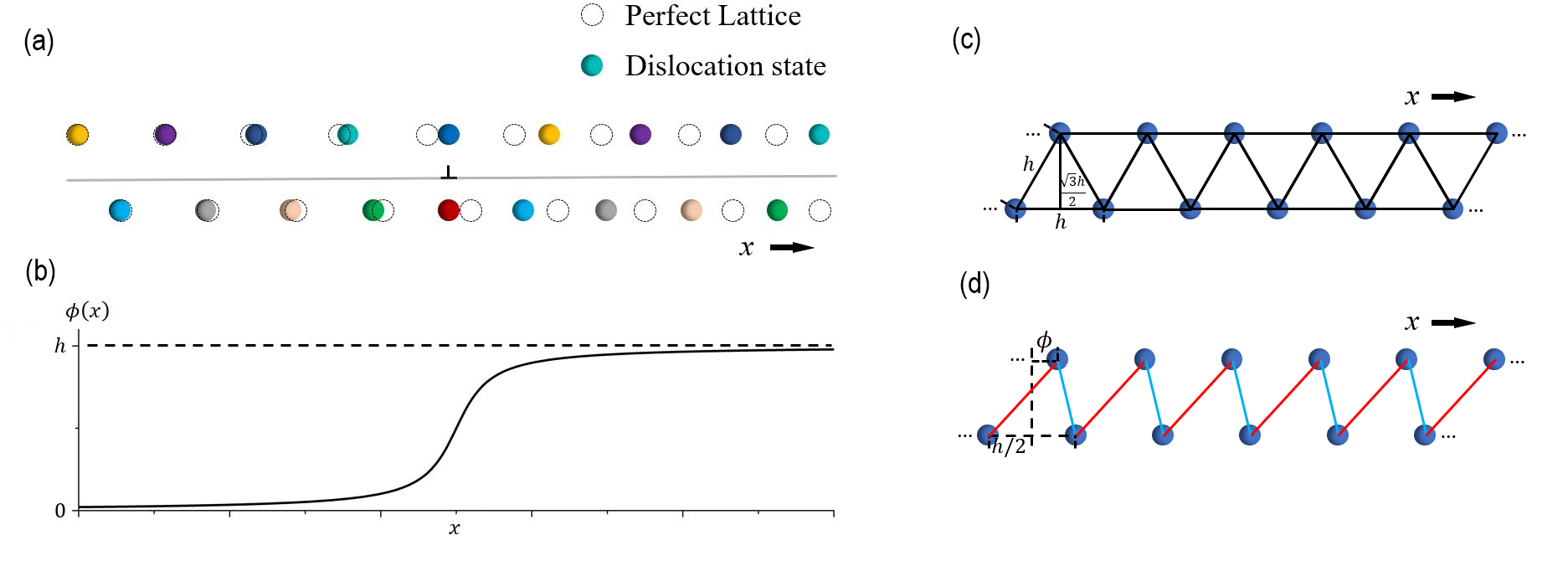}
    \caption{(a) Schematic illustration of the  atomic configuration of an edge dislocation in a bilayer HEA. Different colors meaning different elements. The symbol $\perp$ indicates the location of the dislocation. The light horizontal line represents the slip plane of the dislocation. (b)  Schematic illustration of the average profile of the disregistry $\phi$ for this edge dislocation. (c) The averaged perfect lattice structure (without the dislocation) of a bilayer HEA. The lattice constant is $h$. (d) The lattice with a uniform disregistry of $ \phi $. Red and blue bonds connect first nearest neighbors of inter-layer atoms.  The value of the generalized stacking fault energy $\gamma(\phi)$  is  the energy increment per unit length of this configuration with respect to the perfect lattice  \cite{vitek}.  }
    \label{latticeDislocation}
\end{figure}

HEAs are different from conventional alloys in the sense that each lattice site is randomly occupied by one of the main elements (normally more than five) with nearly equal proportions.
We focus on  a bilayer HEA with an inter-layer straight edge dislocation; see Fig.~\ref{latticeDislocation}(a) for an illustration of the atomic configuration (to be explained at the end of this section).
The averaged perfect lattice structure (without dislocation)  has a triangular atomic configuration, see Fig.~\ref{latticeDislocation}(c).

%\begin{figure}[htbp]
%\centering
%    \includegraphics[width=0.7\linewidth]{latticeDislocation.jpg}
%    \caption{(a) Schematic illustration of the  atomic configuration of an edge dislocation in a bilayer HEA. Different colors meaning different elements. (b)  Schematic illustration of the average profile of the disregistry $\phi$ for this edge dislocation.}
%    \label{latticeDislocation}
%\end{figure}
%
%
%\begin{figure} [tbhp]
%	\centering
%	\includegraphics[width=0.55\columnwidth]{lattice.png}
%	\caption{\label{lattice}(a) The averaged perfect lattice structure of a bilayer HEA. The lattice constant is $h$. (b) The lattice with disregistry of $ \phi $. Red and blue lines connect first nearest neighbors of inter-layer atoms. }
%\end{figure}

The randomness of lattice occupation is expressed by a  probability model. Assume that  in the bilayer HEA, there are $ m $ elements that could possibly occupy each lattice site. All these elements form a sample space of a random variable $\omega$:
\begin{equation}\label{eqn:omega}
\Omega = \left\{  e_1, e_2, \cdots, e_m  \right\},
\end{equation}
which is equipped with probability measure:
\begin{flalign}
P(e_1) = p_1, P(e_2) = p_2,  \cdots, P(e_m) = p_m, \\
{\rm with}\ p_i\geq 0, \ i = 1,2 \cdots, m, \ {\rm and} \  \sum_{i=1}^{m} p_i = 1. \label{eqn:p}
\end{flalign}
The probability of each element occupying a lattice site is the proportion of this element over all elements in the HEA. Especially, in equimolar HEAs, the probabilities of all elements are equal, i.e., $ p_1 = p_2 \cdots = p_m = 1/m $. At each lattice site, say atom $i$, there is a random variable $\omega_i$ that describes the element on that site. In this paper, we assume that all the random variables $\{\omega_i\}$ for all the lattice sites of the HEA are independent and identically distributed with distribution given in Eqs.~\eqref{eqn:omega}--\eqref{eqn:p}.

 We use pair potential $ V^{\rm pair}(r,\omega_{i_1},\omega_{i_2}) $ in the atomistic model of the HEA, from which the continuum model will be derived. This interatomic potential is  a function of not only inter-atomic distance $ r $ but also the two-side atom species $\omega_{i_1}=\chi_1$ and $\omega_{i_2}=\chi_2$ with $\chi_1, \chi_2\in \Omega $.  We focus on the nearest neighbor interaction in the derivation. An example of such a pair potential is the Lennard-Jones potential \cite{Lennard_jones1}
\begin{equation} \label{Lennard_jones}
 	V^{\rm LJ} (r,\chi_1,\chi_2) = 4\epsilon(\chi_1,\chi_2)\cdot \left( \left( \frac{a(\chi_1,\chi_2)}{r} \right)^{12} - \left( \frac{a(\chi_1,\chi_2)}{r} \right)^{6} \right),
\end{equation}
with Lorentz-Berthelot's combining rules \cite{combine_rule1,combine_rule2}
\begin{flalign}
\epsilon(\chi_1,\chi_2) = \sqrt{ \epsilon(\chi_1,\chi_1)\cdot \epsilon(\chi_2,\chi_2) }, \ \ \
a(\chi_1,\chi_2) = \frac{a(\chi_1,\chi_1) + a(\chi_2,\chi_2)}{2}.
\end{flalign}
 That is, in this potential, the dependence on atom species is defined through the empirical parameters $ \epsilon(\chi_1,\chi_2) $ and $ a(\chi_1,\chi_2) $.
 This  and similar forms  of the Lennard-Jones potential have been used for atomistic simulations of HEAs \cite{sharma2016,caro2015,yen2020} and other systems \cite{LJ_parameter2} in the literature. In the numerical validation after the continuum model is derived, without lose of generality,  we
 will use this  Lennard-Jones potential. Note that this specific potential is only for numerical validation, and the obtained continuum model does not depend on the specific form of the pair potential $ V^{\rm pair}(r,\omega_{i_1},\omega_{i_2})$.

The empirical parameters $ \epsilon $ and $ a $ of the Lennard-Jones potential for some transition metal elements, which are some commonly used ingredients of HEAs, are listed in Table~\ref{table_para} (from \cite{caro2015,LJ_parameter2}).

\begin{table} [tbhp]
	\footnotesize
	\caption{The empirical parameters of Lennard-Jones potential for some transition metals.} \label{table_para}
	\begin{center}
	\begin{tabular} {|c|c|c|c|} \hline
		$ \chi $ & $ a(\chi,\chi) \ (\mathring{A}) $ & $ \epsilon(\chi,\chi) \ (eV) $ & atom radius $ (\mathring{A}) $ \\ \hline
		Cr & 2.336 & 0.502 & 1.66 \\
		
		Co & 2.284 & 0.516 & 1.52 \\
		
		Fe & 2.321 & 0.527 & 1.56 \\
		
		Ni & 2.282 & 0.520 & 1.49 \\
		
		Cu & 2.338 & 0.409 & 1.45 \\
		\hline
	\end{tabular}
	\end{center}
\end{table}

Fig.~\ref{latticeDislocation}(a) shows a schematic illustration of the  atomic configuration of an edge dislocation in a bilayer HEA. Here the length of the Burgers vector $b=h$, where $h$ is the lattice constant. Here the disregistry function across the slip plane $\phi(x)=u^+(x)-u^-(x)$ is defined on discrete lattice sites, with $\bar{\phi}(-\infty)=0$ and $\bar{\phi}(+\infty)=h$, where $\bar{\phi}(x)$ is the averaged value of $\phi(x)$. We will derive a continuum model from this atomistic model in the following sections.

%We generalize the stochastic PN model for HEAs by considering a bilayer atomic system with an edge dislocation. The total energy of the system consists of misfit energy between the two atomic layers and elastic energy associating with each layer. Both the misfit energy and elastic energy are evaluated by counting interatomic potentials between atom pairs.

\section{Stochastic misfit energy of HEAs} \label{section_misfit}
%%%%%%%%%%%%%%%%%%%%%%%%%%%%%%%%%%%%%%%%%%%%%%%%%%%%%%%%%%%%%
%The classical PN model considers the misfit energy as an integral of sinusoidal-type energy density along the slip plane \cite{PN1,PN2}. The sinusoidal form is simply an approximation to account for the fact that energy attains its minimum for perfect lattices \cite{sinusoidal,dislocation_book}. The expression of energy density can be more proper and realistic by considering V{\'i}tek's intrinsic stacking fault energy. However, V{\'i}tek's intrinsic stacking fault only works for single-element crystals, and hence is not applicable to HEAs.
In this section, we first calculate the misfit energy density, i.e., the generalized stacking fault energy of the bilayer HEA using the atomistic model with randomness described in the previous section, and then derive stochastic continuum formulations for the generalized stacking fault energy and the misfit energy.

\subsection{Review of the definition of the generalized stacking fault energy~\cite{vitek}}\label{subsec:gammasurface}
In the definition  proposed by Vitek~\cite{vitek}, for a given plane, the generalized stacking fault energy (or the generalized stacking fault energy) as a function of disregistry $\phi$ is the energy increment per unit area after a perfect crystal is cut along this plane and then reconnected after a uniform shift $\phi$.

%a bulk crystal is separated into two parts by cutting on the crystal plane. One part shifts a fault vector relative to the other part, forming a lattice mismatch between the two parts. The generalized stacking fault energy (or so called $ \gamma $-surface) is the difference of energy density upon the slip plane between before and after fault shifting.
%V{\'i}tek's generalization of stacking fault energy is a function of the 2-dimensional fault vector.
%To apply in PN model, the concept is simplified by replacing the bulk crystal with a 2-dimensional lattice and cutting on a crystal line. Then the fault vector is reduced to be a 1-dimensional disregistry and the energy density becomes a function of the disregistry.

For a bilayer single-element crystal with triangular lattice as shown in Fig.~\ref{latticeDislocation}(c), the generalized stacking fault energy $\gamma(\phi)$ is the energy increment per unit length after the top and bottom layers have a uniform shift (disregistry) $\phi$ relative to each other (i.e. along the $x$ direction); see Fig.~\ref{latticeDislocation}(d). This is the traditional crystal and can be regarded as a special case under our framework when there is only one possible element in the probability space, i.e. $ \Omega = \{ e_1 \} $ with $ P(e_1) = 1 $.
In this classical case, the interatomic potential becomes a function of only distance $ r $. When the nearest neighbor interaction is considered, under a uniform inter-layer disregistry $\phi$,  the increment in the interaction energy of one atom with all the other atoms  consists of  increments of the interaction energies with two nearest neighbors on the other layer $ U(\phi) $ and $ V(\phi) $ (due to the red and blue bonds, respectively, in Fig.~\ref{latticeDislocation}(d)):
\begin{subequations} \label{fnn}
	\begin{align}
		U(\phi) &= V^{\rm pair}\left( \sqrt{\Big( \frac{h}{2} + \phi \Big)^2 + \Big( \frac{\sqrt{3}h}{2} \Big)^2}, \chi_1, \chi_1 \right) - V^{\rm pair}\Big(h,\chi_1,\chi_1\Big), \\
%		&= V^{\rm pair}\Big( \sqrt{\phi^2 + h\phi + h^2}, \chi_1, \chi_1 \Big) - V^{\rm pair}\Big(h,\chi_1,\chi_1\Big),\nonumber\\
		V(\phi) &= V^{\rm pair}\left( \sqrt{\Big( \frac{h}{2} - \phi \Big)^2 + \Big( \frac{\sqrt{3}h}{2} \Big)^2}, \chi_1, \chi_1 \right) - V^{\rm pair}\Big(h,\chi_1,\chi_1\Big).
%	\\	&= V^{\rm pair}\Big( \sqrt{\phi^2 - h\phi + h^2}, \chi_1, \chi_1 \Big) - V^{\rm pair}\Big(h,\chi_1,\chi_1\Big). \nonumber
\end{align}
\end{subequations}
Here we have used the fact that with the uniform inter-layer disregistry $\phi$, the distances between one atom with its two nearest neighbors in the same layer do not change, thus the associated interaction energies do not change and do not contribute to the energy increment.
Therefore, the generalized stacking fault energy $ \gamma(\phi) $ can be expressed as
\begin{equation} \label{gamma_vitek}
	\gamma(\phi) = \dfrac{1}{h} \Big[ U(\phi) + V(\phi) \Big].
\end{equation}

\begin{figure} [tbhp]
	\centering
	\includegraphics[width=0.5\columnwidth]{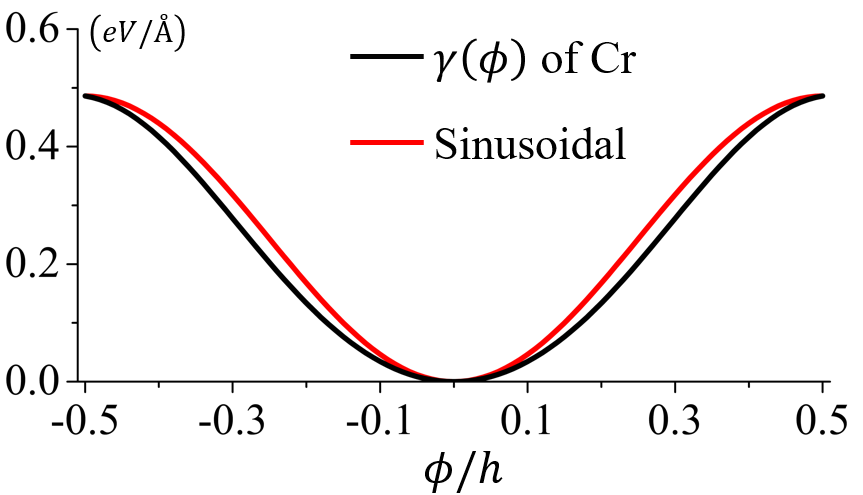}
	\caption{\label{gammaCr} The black line is $ \gamma(\phi) $ of Chromium, and the red line is the fitting sinusoidal curve.}
\end{figure}

Fig.~\ref{gammaCr} shows an example of $\gamma(\phi)$, calculated using the
 parameters of Chromium from Table~\ref{table_para}. Note that $ \gamma(\phi) $ is a periodic function with  period of $ h $. The approximation of the Frenkel's sinusoidal-type potential in Eq.~\eqref{eqn:Frenkel} \cite{Frenkel1926} adopted in the classical Peierls-Nabarro model \cite{PN1,PN2} is also plotted in Fig.~\ref{gammaCr}, with the same period and amplitude as the calculated $\gamma(\phi)$. It can be seen that the  Frenkel  sinusoidal potential indeed provides a good approximation to the generalized stacking fault energy in this case. This also validates that using the Frenkel  sinusoidal potential as the averaged nonlinear inter-layer potential in the studies of HEAs in Ref.~\cite{theoretical1} is a reasonable approximation.

\subsection{Stochastic generalized stacking fault energy}\label{subsec:gsf}
\begin{figure} [tbhp]
	\centering
	\includegraphics[width=0.55\columnwidth]{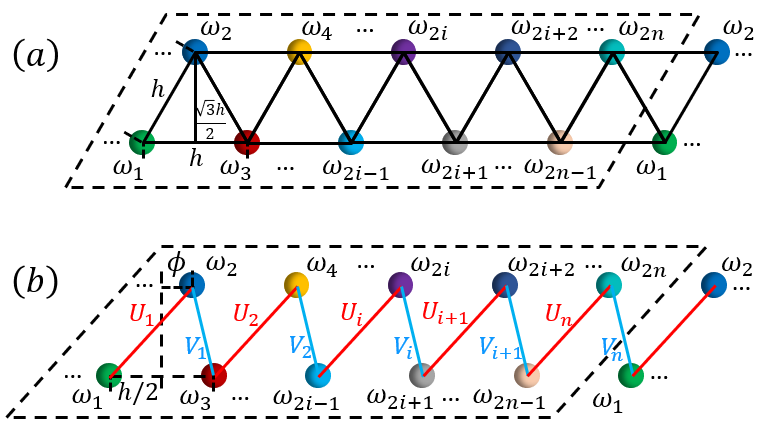}
	\caption{\label{latticeMisfit} (a) The dashed box is a supercell of type-$ n $ in the bilayer HEA with triangular lattice structure.  Different colors denote different atom species. Periodic boundary condition is adopted for the supercell. (b)  Supercell with disregistry of $ \phi $. Red and blue bonds connect the nearest neighbors of inter-layer atoms.}
\end{figure}

In order to incorporate the atomic level randomness into the continuum model, we introduce the concept of supercell. One supercell of type-$ n $ contains $ 2n $ atoms ($n$ atoms on each layer), with species denoted by random variables $ \omega_1, \omega_2, \cdots ,\omega_{2n} \in \Omega $. We further define the atomic configuration of the supercell as $ \omega:= (\omega_1,\omega_2,\cdots, \omega_{2n}) \in\Omega^{2n} $. Periodic boundary condition is used for the supercell. See Fig.~\ref{latticeMisfit} for illustrations of the supercell and supercell with a disregistry $ \phi $ for the calculation of the generalized stacking fault energy. The atoms in the upper layer are labeled as $2i$, $i=1,2, \cdots, n$, and those in the lower layer are $2i-1$, $i=1,2, \cdots, n$.

We will derive a continuum model from the atomistic model under the assumption that the size of the supercell $\delta=nh$ is large on the atomic level and small on the continuum level, i.e., $h \ll \delta \ll L$, where $L$ is the length scale of the continuum model. The derivation will be given in Secs.~\ref{eqn:continuum-gamma} and \ref{subsec:convergenceproof}.
%Here we obtain the continuum model by asymptotic analysis and numerical samplings. More rigorous proof using a modified central limit theorem will be given in Sec.~\ref{subsec:convergenceproof}, together with the convergence of the stochastic elastic energy.

%In Figure~\ref{latticeMisfit}, the dashed box is one supercell of type-$ n $, where different atom species are represented by different colors. One such supercell repeats infinite times forming a bilayer system.
%Figure~\ref{latticeMisfit}(a) illustrates the equilibrium state of the bilayer system with lattice constant $ h $.
%Figure~\ref{latticeMisfit}(b) illustrates the same bilayer system with top layer relatively shifting a fault disregistry $ \phi $.
%One supercell of type-$ n $ contains $ 2n $ atoms in total, with species of bottom layer denoted by $ \omega_1, \omega_3, \cdots , \omega_{2n-1} \in \Omega $ and species of top layer denoted by $ \omega_2, \omega_4, \cdots , \omega_{2n} \in \Omega $. We further define the atomic configuration of the supercell as $ \omega \coloneqq (\omega_1,\omega_2,\cdots,\omega_{2n}) \in \Omega^{2n} $. We also note here that the periodicity of the bilayer system indicates that $ \omega_{2n+i} = \omega_i $ for integers $ i = 1,2,\cdots,2n $.

We have assumed that the occupation of atom species on one lattice site is independent from that of any other sites. Therefore, the probability measure of any atomic configuration is well established by direct product of the probability from each site:
\begin{equation} \label{measureomega}
	P(\omega) = P(\omega_1) \cdot P(\omega_2) \cdots P(\omega_{2n}) = \prod_{i=1}^{2n} P(\omega_i).
\end{equation}
The interaction energy between each pair of atoms and the total interaction energy within the supercell are functions of  $ \omega $.

%We evaluate stacking fault energy by accounting pair potentials between inter-layer atoms. Because $ V^{pair}(r,\chi_1,\chi_2) $ is defined as a function of two-side atom species, the pair potentials within the supercell all become functions of $ \omega $.

With a uniform inter-layer disregistry $\phi$, following the formulation of deterministic case in Eq.~\eqref{fnn}, the increment in the
 interaction energy of one atom, without loss of generality, the atom with label $2i$ in the upper layer, with all the other atoms consists of the interaction energy increments of atom $2i$ with its two nearest neighbors $2i-1$ and $2i+1$ in the lower layer:
\begin{subequations} \label{def_uv}
	\begin{align}
	U_i(\phi,\omega) &= V^{\rm pair}\left( \sqrt{\phi^2 + h\phi + h^2}, \omega_{2i-1},\omega_{2i} \right) - V^{\rm pair}(h,\omega_{2i-1},\omega_{2i}), \\
	V_i(\phi,\omega) &= V^{\rm pair}\left( \sqrt{\phi^2 - h\phi + h^2}, \omega_{2i},\omega_{2i+1} \right) - V^{\rm pair}(h,\omega_{2i},\omega_{2i+1}).
	\end{align}
\end{subequations}
 Therefore, for this type-$ n $  supercell  with atomic configuration $ \omega $ under  disregistry $ \phi $, the value of the generalized stacking fault energy, i.e., the average energy increment per unit length of the supercell, is
\begin{equation} \label{gamma_n}
	\gamma_n(\phi,\omega) = \dfrac{1}{nh}\sum_{i=1}^{n} \Big[ U_i(\phi,\omega) + V_i(\phi,\omega) \Big].
\end{equation}

If the probability space contains only one possible element, i.e. $ \Omega = \{ e_1 \} $, each lattice site should be occupied by this element with probability 1. In this extreme case, the stochastic $ \gamma_n $ in Eq.~\eqref{gamma_n} reduces to be the classical, deterministic expression in Eq.~\eqref{gamma_vitek}. This indicates that our definition of stochastic generalized stacking fault energy is consistent with the classical definition by Vitek \cite{vitek}.

Now we
 calculate the mean and variance of $ \gamma_n(\phi,\omega) $.
Since the random variables $\omega_i$ for the elements on the lattice sites have identical distribution,
using Eq.~\eqref{gamma_n}, the mean of $ \gamma_n (\phi,\omega) $ is
\begin{equation} \label{gammaexpectation}
	\bar{\gamma}(\phi) := \mathbb{E} \Big[ \gamma_n(\phi,\omega) \Big] = \dfrac{1}{h} \Big[ \bar{U}(\phi) + \bar{V}(\phi) \Big],
\end{equation}
where
\begin{equation}\label{eqn:uvbar}
		\bar{U} (\phi) := \mathbb{E} \Big[ U_i(\phi,\omega) \Big], \ \ \
		\bar{V} (\phi) := \mathbb{E} \Big[ V_i(\phi,\omega) \Big].
\end{equation}
for $i=1,2,\cdots,n$.

%\begin{align} \label{gammaexpectation}
%	\bar{\gamma}(\phi) :=& \mathbb{E} \Big[ \gamma_n(\phi,\omega) \Big] = \dfrac{1}{nh} \sum_{i=1}^{n} \mathbb{E} \Big[ U_i(\phi,\omega) + V_i(\phi,\omega) \Big]  \nonumber \\
%	=& \dfrac{1}{h} \Big[ \bar{U}(\phi) + \bar{V}(\phi) \Big].
%\end{align}

Next, we calculate the variance of $ \gamma_n (\phi,\omega) $.  Subtracting (\ref{gammaexpectation}) from (\ref{gamma_n}), we have
\begin{equation}\label{eqn:subtract}
	 \gamma_n(\phi,\omega) =  \bar{\gamma}(\phi) + \frac{1}{nh}\sum_{i=1}^{n} \Big[ (U_i - \bar{U}) + (V_i - \bar{V}) \Big].
\end{equation}
The variance of $ \gamma_n (\phi,\omega) $ is
\begin{equation}
{\rm Var}\big(\gamma_n (\phi,\omega)\big)=\mathbb{E} \Big[\big(\gamma_n (\phi,\omega)-\bar{\gamma}(\phi)\big)^2\Big]=\frac{1}{n^2h^2}\mathbb{E} \Big[\Big( \sum_{i=1}^{n} \big[ (U_i - \bar{U}) + (V_i - \bar{V}) \big] \Big)^2\Big].
\end{equation}

It can be calculated that
\begin{flalign}\label{sigma1}
	&\mathbb{E}\Big[ \Big( \sum_{i=1}^{n} \big[ (U_i - \bar{U}) + (V_i - \bar{V}) \big] \Big)^2  \Big] \\
		=& \mathbb{E} \Big[ \sum_{i=1}^{n} ( U_i - \bar{U}) \Big]^2 + \mathbb{E} \Big[ \sum_{i=1}^{n} ( V_i - \bar{V}) \Big]^2 + 2 \cdot \mathbb{E}\Big[ \sum_{i,j=1}^{n} (U_i - \bar{U})(V_j - \bar{V}) \Big]  \nonumber \\
		=& \sum_{i=1}^{n} \mathbb{E}\Big[ (U_i - \bar{U})^2\Big] + \sum_{i=1}^{n} \mathbb{E}\Big[ (V_i - \bar{V})^2\Big] + 2\cdot\sum_{i=1}^{n} \mathbb{E} \Big[ (U_i - \bar{U})(V_i + V_{i-1} - 2\bar{V}) \Big]   \nonumber  \\
		=& n\sigma_{uu} + n\sigma_{vv} + 2n\sigma_{uv},  \nonumber
	\end{flalign}
where
\begin{subequations}
	\begin{align}
		&\sigma_{uu}(\phi) := \mathbb{E}\Big[(U_i - \bar{U})^2\Big], \ \
		\sigma_{vv}(\phi) := \mathbb{E}\Big[(V_i - \bar{V})^2\Big], \label{sigma_uu} \\
		&\sigma_{uv}(\phi) := \mathbb{E} \Big[ (U_i - \bar{U})(V_i + V_{i-1} - 2\bar{V}) \Big]. \label{sigma_uv}
	\end{align}
\end{subequations}
Here we have used the fact that all $ U_i $ and all $ V_i $ have identical distributions, respectively. Moreover, since the random variables $\omega_i$ for the elements on the lattice sites are independent to each other, each  $ U_i $ is correlated only with  $ V_{i-1} $ and $ V_i $ and is independent with all the other $ V_j$'s, see  Fig.~\ref{latticeMisfit}(b). This leads to
$\mathbb{E}\Big[ \sum_{i,j=1}^{n} (U_i - \bar{U})(V_j - \bar{V}) \Big] =\sum_{i=1}^{n} \mathbb{E} \Big[ (U_i - \bar{U})(V_i + V_{i-1} - 2\bar{V}) \Big]$
in the above equations.

Introducing the notation $ \theta(\phi) $:
\begin{equation} \label{theta}
\theta(\phi)  := \sqrt{\big[\sigma_{uu}(\phi) + \sigma_{vv}(\phi) + 2\sigma_{uv}(\phi) \big]/h} \ ,
\end{equation}
we have
\begin{equation} \label{var_msf}
	\mathbb{E} \Big( \sum_{i=1}^{n} \big[ (U_i - \bar{U}) + (V_i - \bar{V}) \big] \Big)^2 = nh\cdot \theta^2(\phi),
\end{equation}
\begin{equation}\label{eqn:variance-gamma}
	{\rm Var}\big(\gamma_n (\phi,\omega)\big) = \frac{1}{nh}\cdot \theta^2(\phi).
\end{equation}

\subsection{Continuum limit of the stochastic generalized stacking fault energy}\label{eqn:continuum-gamma}
In this subsection, we will derive a continuum formulation of $\gamma(\phi,\omega)$
from the atomic-level expression $ \gamma_n(\phi,\omega) $  in Eq.~\eqref{gamma_n}
 by letting the size of the supercell $n\rightarrow\infty$.
  Here we perform numerical samplings to examine this limit. More rigorous convergence proof using a modified central limit theorem will be given in Sec.~\ref{subsec:convergenceproof}.

\begin{figure} [tbhp]
	\centering
	\includegraphics[width=0.5\columnwidth]{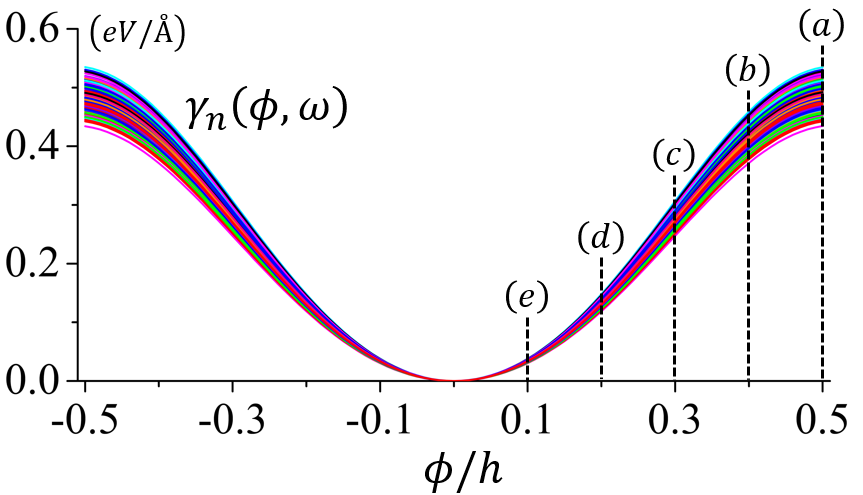}
	\caption{\label{gammamulti} Total number of $ 10^6 $ sampling curves of $ \gamma_n(\phi,\omega) $, where the size of the supercell is $n=7$. The vertical dash lines denotes the five fixed values of disregistry: (a) $ \phi = 0.5h $, (b) $ \phi = 0.4h $, (c) $ \phi = 0.3h $, (d) $ \phi = 0.2h $ and (e) $ \phi = 0.1h $.}
\end{figure}

 We consider an HEA that consists of the five elements shown in Table~\ref{table_para}. The probability space equipped with probability measure is
\begin{equation} \label{example}
	\begin{aligned}
		\Omega &= \left\{ \text{Cr, Co, Fe, Ni, Cu} \right\}, \\
		P(\text{Cr}) = P(&\text{Co}) = P(\text{Fe}) = P(\text{Ni}) = P(\text{Cu}) = 1/5.
	\end{aligned}
\end{equation}
In this calculation example, we set one supercell containing $n=7$ atom pairs. We sample total number of $ 10^6 $ atomic configurations by the probability distribution \eqref{measureomega}. Each atomic configuration $ \omega_{\rm sample} $ corresponds to one curve of $ \gamma_n(\phi,\omega_{\rm sample}) $ shown in Fig.~\ref{gammamulti}.

\begin{figure} [tbhp]
	\centering
	\includegraphics[width=0.6\columnwidth]{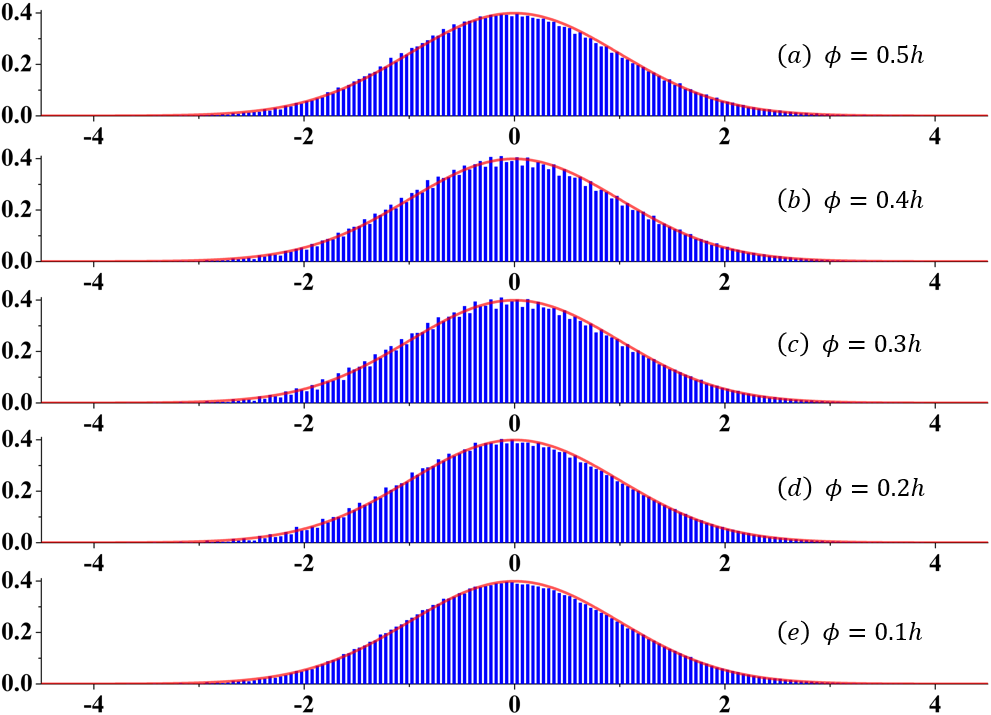}
	\caption{\label{gammadistribution} Probability distributions of $ \gamma_n(\phi,\omega) $ at five fixed disregistry values, normalized by  the mean and variance  in Eq.~\eqref{gammaexpectation}  and \eqref{eqn:variance-gamma}. (a) $ \phi = 0.5 h $, (b) $ \phi = 0.4 h $, (c) $ \phi = 0.3 h $, (d) $ \phi = 0.2 h $, (e) $ \phi = 0.1 h $. Bar graphs are normalized distributions of $ 10^6 $ samples. The red lines are probability density functions of the Gaussian distributions with mean $0$ and standard deviation $1$.}
\end{figure}

With all those samples, we also statistically find the distributions of $ \gamma $-values at five fixed disregistry, namely $ \phi = 0.5 h $, $ 0.4 h $, $ 0.3 h $, $ 0.2 h $ and $ 0.1 h $. The total $ 10^6 $ samples indicates there are $ 10^6 $ sampling of values of $ \gamma_n(\phi,\omega) $  in Eq.~\eqref{gamma_n} at each fixed disregistry $ \phi$. Fig.~\ref{gammadistribution} shows the normalized distributions of those sampling $ \gamma $-values at each of these values of  $ \phi$, using the mean and variance of $ \gamma_n(\phi,\omega) $ in Eq.~\eqref{gammaexpectation}  and \eqref{eqn:variance-gamma}, and comparison with the probability density function of Gaussian distribution with mean $0$ and standard deviation $1$.
  The results shows that the sample distributions agree excellently with the Gaussian distributions for this supercell with size $n=7$. We have also performed samplings with larger sizes of the supercell, and the results are almost identical to those shown in Figs.~\ref{gammamulti} and \ref{gammadistribution}.

These numerical results show that for each value of the disregistry $\phi$, the value of the generalized stacking fault energy $\gamma_n(\phi,\omega)$ converges to a random variable with Gaussian distribution. That is
\begin{equation} \label{statement1-1}
	\dfrac{\sqrt{nh}}{\theta(\phi)} \ \big[\gamma_n (\phi,\omega)-\bar{\gamma}(\phi)\big]   \longrightarrow \mathcal{N}(0,1), \ \ \  \text{as} \ \ n \longrightarrow \infty,
\end{equation}
where $\mathcal{N}(\mu,\sigma^2)$ is the Gaussian distribution with mean $\mu$ and standard deviation $\sigma$.
The numerical results show that the convergence is already quite good for $n=7$. More rigorous convergence proof using a modified central limit theorem will be given in Sec.~\ref{subsec:convergenceproof}.

The above limit is equivalent to
\begin{equation} \label{statement1}
	\dfrac{1}{\sqrt{nh}\cdot \theta(\phi)} \sum_{i=1}^{n} \Big[ (U_i - \bar{U}) + (V_i - \bar{V}) \Big] \longrightarrow \mathcal{N}(0,1), \ \ \  \text{as} \ \ n \longrightarrow \infty,
\end{equation}
which will be used in later derivation.

%Although the statement (\ref{statement1}) requires $ n \longrightarrow \infty $, the calculation results Figure~\ref{gammadistribution} of the example illustrates that when $ n $ is even less than 10, the $ \gamma $-values can be well approximated by normal distributions.

\subsection{Stochastic misfit energy}\label{sec:stochastic_misfit}
Now we derive the formulation for the misfit energy based on the stochastic generalized stacking fault energy $\gamma_n(\phi,\omega)$.

We have assumed that the size of the supercell $ \delta = nh $ is much smaller than the length unit of the continuum model. Defining
\begin{equation}\label{eqn:y_delta}
Y_{\delta}(\omega) \sim \mathcal{N}(0,\delta),
\end{equation}
 which is the Gaussian distribution with mean $0$ and standard deviation $\sqrt{\delta}$, and using Eq.~\eqref{eqn:subtract} and \eqref{statement1}, the misfit energy within the supercell is
\begin{align} \label{atomic_msf}
	\Delta E_{\rm misfit} &= nh\cdot \gamma_n(\phi,\omega)  \\
	&= nh \cdot \bar{\gamma}(\phi) + \sum_{i=1}^{n} \Big[ (U_i - \bar{U}) + (V_i - \bar{V}) \Big] \nonumber\\
	&\cong nh \cdot \bar{\gamma}(\phi) + \sqrt{nh} \cdot \theta(\phi) \cdot Y_1 \nonumber\\
	&= \delta\cdot\bar{\gamma}(\phi) + \theta(\phi)\cdot Y_{\delta}. \nonumber
\end{align}

We discretize the slip plane $x$-axis into a series of such small intervals meaning microscopic supercells: $\delta_1,\delta_2, \delta_3, \cdots $, and each interval is associated with a Gaussian random variable for the atomic structure within it: $ Y_{\delta_1}, Y_{\delta_2}, Y_{\delta_3}, \cdots $. (For the infinite domain, we can start from a finite number $A<0$ and then let $A\rightarrow -\infty$.) Because the atomic configuration within one interval is almost independent from that of any other interval due to the assumptions of nearest neighbor interaction and $\delta\gg h$, $ Y_{\delta} $'s are approximately mutually independent and can be regarded as independent Gaussian increments. Therefore, the sequence of $ \{Y_{\delta}\} $ defines a Brownian motion (Wiener process)
 $ B_x(\omega) $ as
\begin{equation} \label{brownian_motion}
	Y_{\delta} = B_{x + \delta}(\omega) - B_x(\omega).
\end{equation}
Since $ \delta $ is small on the continuum length scale, the microscopic misfit energy in Eq.~\eqref{atomic_msf} can be written on the continuum length scale as
\begin{equation} \label{diff_msf}
	d E_{\rm misfit} = \bar{\gamma}(\phi) dx + \theta(\phi) d B_x.
\end{equation}
This is the formulation of the stochastic misfit energy on the continuum level.

In the extreme case that there is only one possible element in the probability space, i.e. $ \Omega = \{ e_1 \} $ with $ P(e_1) = 1 $, then $ \theta(\phi) \equiv 0 $ and the formulation of the misfit energy reduces to that in the  classical Peierls-Nabarro model shown in Eq.~\eqref{eqn:Emisfit}.

\section{Stochastic elastic energy} \label{section_elstic}
In this section, we first calculate the energy due to the intra-layer elastic interaction of the bilayer HEA using the atomistic model,  and then derive stochastic continuum formulation from it.

\subsection{Elastic energy using the atomistic model}
\begin{figure} [tbhp]
	\centering
	\includegraphics[width=0.7\columnwidth]{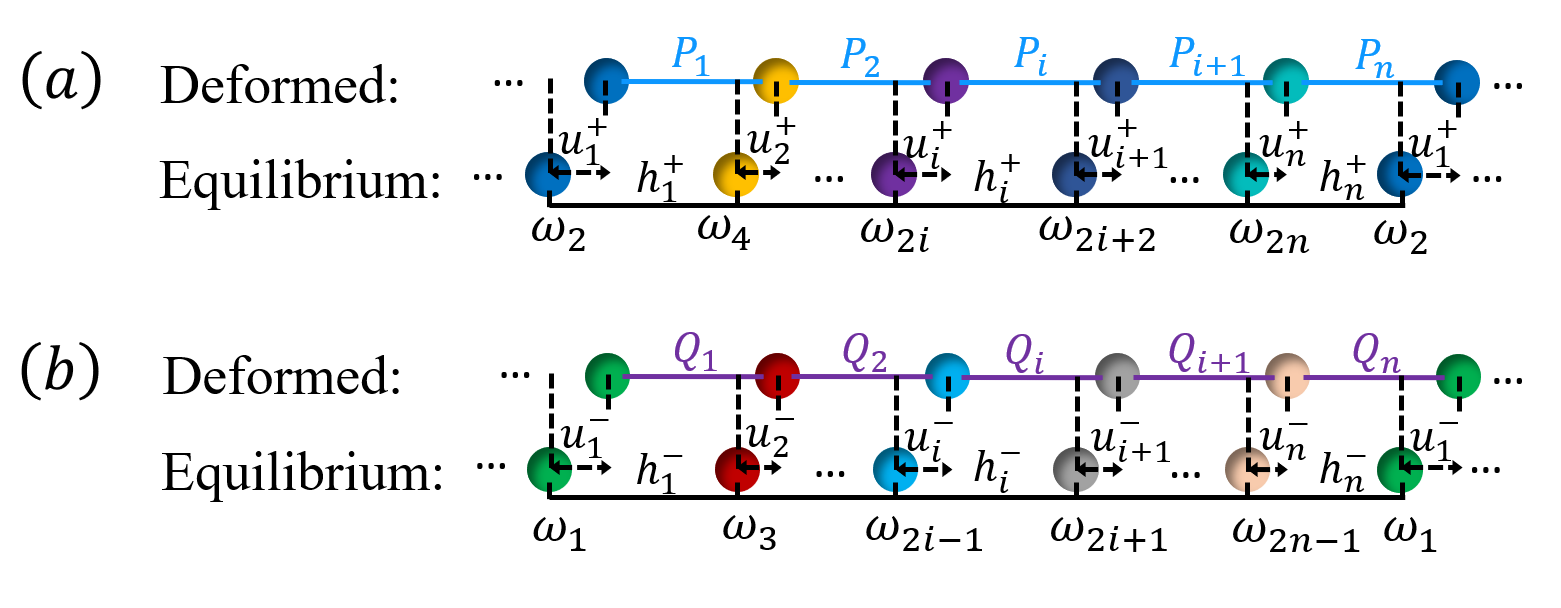}
	\caption{\label{latticeElastic} Top (a) and  bottom (b) layers of the supercell with and without atom displacements. Blue and purple lines connect the interacting  nearest intra-layer  neighboring atoms. The displacements of the $ i $'th atoms of the top and bottom layers are $ u_i^+ $ and  $ u_i^- $, respectively. $P_i$ and $Q_i$ are the elastic energy stored in the bond between the $ i $'th and the $ (i + 1) $'th atoms in the top and bottom layers, respectively. }
\end{figure}

The elastic energy comes from the pairwise interactions  between intra-layer neighboring atoms. Fig.~\ref{latticeElastic} illustrates one supercell with and without displacements. The supercell for calculating the elastic energy is the same as that for evaluating the misfit energy, i.e. the atom configuration $ \omega = (\omega_1,\omega_2,\cdots,\omega_{2n}) $ in Fig.~\ref{latticeElastic} is the same as that in Fig.~\ref{latticeMisfit}.  We set the displacement of the $ i $'th atom of the top layer as $ u_i^+ $, and that of the bottom layer as $ u_i^- $.

%In the following content of this subsection, we first figure out the equilibrium state of the supercell by calculating equilibrium distance of each nearest-neighbor pair. Then in the state with displacements, the elastic energy of each nearest-neighbor pair is approximated by expanding their potential to the second order. The total elastic energy of the supercell is the summation of elastic energy from each pair.

Because only nearest-neighbor interactions are considered in our model, the equilibrium  atomic lattice is reached when the distance between each nearest neighbors is the energy minimum distance of the pair potential. In fact, in this case, the total energy of the lattice is minimized, as can be seen from the fact that any perturbation of the location of an atom will lead to increase of the total energy.
%When lattice is in equilibrium state, its total energy takes minimum. The total energy consists of only nearest-neighbor potentials because only nearest-neighbor interactions are considered in our model. Therefore, it is sufficient to consider nearest-neighbor potential being minimum in equilibrium.

We consider a nearest-neighbor pair with species generally noted as $ \chi_1, \chi_2 $ whose interaction is given by the  pair potential $ V^{\rm pair}(r,\chi_1,\chi_2) $, where $r$ is the distance between them. The equilibrium distance $ h $ is the value when pair potential reaches minimum, i.e., $ r = h $ is the solution of
\begin{equation} \label{dvdr}
	\dfrac{d V^{\rm pair}}{d r} = 0.
\end{equation}
Because the pair potential $ V^{\rm pair} $ is dependent on atom species $ \chi_1 $ and $ \chi_2 $, the equilibrium distance $h$ determined by solving Eq.~\eqref{dvdr} should also be regarded as a function of the pair species, i.e. $ h = h(\chi_1,\chi_2) $. For the Lennard-Jones potential in Eq.~\eqref{Lennard_jones}, the equilibrium distance $ h(\chi_1,\chi_2) = 2^{1/6}a(\chi_1,\chi_2) $.
For the supercell shown in Fig.~\ref{latticeElastic}, we denote the equilibrium distance of the $ i $'th nearest-neighbor pair as
%\begin{subequations}
	\begin{align}
		h_i^+ := h(\omega_{2i},\omega_{2i+2}), \ \ \
		h_i^- := h(\omega_{2i-1},\omega_{2i+1}),
	\end{align}
%\end{subequations}
for the top  and bottom layers, respectively.

When the atomic lattice is deformed, the elastic energy stored in the bond between the $ i $'th and the $ (i + 1) $'th atoms for the top layer can be expressed as (see Fig.~\ref{latticeElastic}):
\begin{flalign} \label{Pi}
	P_i &= V^{\rm pair}(h_{i}^{+} + u_{i+1}^{+} - u_{i}^{+}, \omega_{2i}, \omega_{2i+2}) - V^{\rm pair}(h_{i}^{+},\omega_{2i},\omega_{2i+2}) \nonumber\\
	&= \dfrac{(h_{i}^{+})^2}{2} \dfrac{d^2 V^{\rm pair}}{d r^2} \Big|_{r = h_{i}^{+}} \cdot \left( \dfrac{u_{i+1}^{+} - u_i^+}{h_i^+} \right)^2 + O(u_{i+1}^+ - u_i^+)^3 \nonumber \\
	&\approx \dfrac{1}{2} \beta_i^+ \left( \dfrac{d u^+(x)}{d x} \right)^2,
\end{flalign}
and for the bottom layer as
\begin{flalign} \label{Qi}
	Q_i &= V^{\rm pair}(h_{i}^{-} + u_{i+1}^{-} - u_{i}^{-}, \omega_{2i-1}, \omega_{2i+1}) - V^{\rm pair}(h_{i}^{-},\omega_{2i-1},\omega_{2i+1}) \nonumber\\
	&= \dfrac{(h_{i}^{-})^2}{2} \dfrac{d^2 V^{\rm pair}}{d r^2} \Big|_{r = h_{i}^{-}} \cdot \left( \dfrac{u_{i+1}^{-} - u_i^-}{h_i^-} \right)^2 + O(u_{i+1}^- - u_i^-)^3 \nonumber \\
	&\approx \dfrac{1}{2} \beta_i^- \left( \dfrac{d u^-(x)}{d x} \right)^2.
\end{flalign}
The variables $ u^+(x) $ and $ u^-(x) $ are notations for the continuous displacements of the top and bottom layers, respectively, and the  stiffness coefficients $\beta_i^+$ and $\beta_i^-$ are defined as
\begin{subequations} \label{beta}
	\begin{align}
		\beta_i^+ = \beta_i^+(\omega_{2i},\omega_{2i+2})   &:= \left( h_i^+ \right)^2 \dfrac{d^2 V^{\rm pair}}{d r^2} \Big|_{r = h_i^+}, \\
		\beta_i^- = \beta_i^-(\omega_{2i-1},\omega_{2i+1}) &:= \left( h_i^- \right)^2 \dfrac{d^2 V^{\rm pair}}{d r^2} \Big|_{r = h_i^-},
	\end{align}
\end{subequations}
for the $ i $'th neighboring atom pairs of the top  and bottom layers, respectively. The elastic energies associated with them, i.e., $P_i$ and $Q_i$ in Eqs.~\eqref{Pi} and \eqref{Qi}, are in the form of Hooke's law. As shown in (\ref{beta}), stiffness coefficients $ \beta_i^{\pm} $ are random variables depending only on the species of the $ i $'th atom neighbor.

As in the previous section, we will derive a continuum model from the atomistic model under the assumption that the size of the supercell $\delta=nh$ is large on the atomic level and small on the continuum level, i.e., $h \ll \delta \ll L$, where $L$ is the length scale of the continuum model.
Following the Cauchy-Born rule \cite{BornHuang1954} for deriving continuum model from the atomistic model for an elastically deformed crystal, we assume that the deformation gradient, which is $\frac{d u^+}{dx}$ or $\frac{d u^-}{dx}$ in the top or bottom layer here, is constant in the supercell.
Under this assumption,
the elastic energy of the supercell for the top or bottom layer, which  is the summation of all the bonding energy of the layer, can be expressed as
\begin{subequations}\label{eqn:elastic0}
	\begin{align}
		\Delta E_{\rm elastic}^+ = \sum_{i=1}^{n} P_i = \dfrac{1}{2} \left(\sum_{i=1}^{n} \beta_i^+\right)\left( \dfrac{d u^+}{dx} \right)^2 , \\
		\Delta E_{\rm elastic}^- = \sum_{i=1}^{n} Q_i = \dfrac{1}{2} \left(\sum_{i=1}^{n} \beta_i^-\right)\left( \dfrac{d u^-}{dx} \right)^2.
	\end{align}
\end{subequations}

\subsection{Mean and variance of the stiffness coefficients}\label{subsec:elastic-mean-var}
The randomness in the elastic energies in Eq.~\eqref{eqn:elastic0} is associated with the random
stiffness coefficients $ \beta_i^{+} $ and $ \beta_i^{-} $ defined in Eq.~\eqref{beta}.
Because the stiffness coefficients $ \beta_i^{+} $ and $ \beta_i^{-} $   are only dependent on the species of the neighboring atoms, their mean values are the same and independent with respect to the index $ i $. The mean value of them is
\begin{equation} \label{mean_beta}
	\bar{\beta} = \mathbb{E} [\beta_i^+] = \mathbb{E} [\beta_i^-].
\end{equation}
Introducing the elastic constant $ \bar{\alpha} $:
\begin{equation} \label{mean_alpha}
	\bar{\alpha} = \bar{\beta} /h,
\end{equation}
the elastic energies in Eq.~\eqref{eqn:elastic0} can be written as
\begin{align} \label{atomic_elastic}
	\Delta E_{\rm elastic}^{\pm} &= \dfrac{1}{2}  \left( \sum_{i=1}^{n} \beta_i^{\pm}\right)\left( \dfrac{d u^{\pm}}{dx} \right)^2
 = \dfrac{1}{2} n \bar{\beta} \left( \dfrac{d u^{\pm}}{dx} \right)^2
 + \dfrac{1}{2}  \left( \sum_{i=1}^{n} \left( \beta_i^{\pm} - \bar{\beta} \right)\right)\left( \dfrac{d u^{\pm}}{dx} \right)^2  \nonumber \\
	&= nh \cdot \dfrac{1}{2} \bar{\alpha} \left( \dfrac{d u^{\pm}}{dx} \right)^2 + \dfrac{1}{2}  \left( \sum_{i=1}^{n} \left( \beta_i^{\pm} - \bar{\beta} \right)\right)\left( \dfrac{d u^{\pm}}{dx} \right)^2.
\end{align}

The randomness in this elastic energy is associated with the random variable $\sum_{i=1}^n \left( \beta_i^\pm - \bar{\beta} \right)$. The mean of  $\sum_{i=1}^n \left( \beta_i^\pm - \bar{\beta} \right)$ is $0$, and its variance is
\begin{equation} \label{varbeta}
	\mathbb{E} \left[ \sum_{i=1}^n \left( \beta_i^+ - \bar{\beta} \right) \right]^2 = \mathbb{E} \left[ \sum_{i=1}^n \left( \beta_i^- - \bar{\beta} \right) \right]^2=n\sigma_{\beta\beta}.
\end{equation}
It can be calculated that
\begin{equation} \label{sigmabeta}
	\sigma_{\beta\beta} = \mathbb{E}\Big[ \left( \beta_i^{\pm} - \bar{\beta} \right) \left( \beta_{i-1}^{\pm} + \beta_i^{\pm} + \beta_{i+1}^{\pm} - 3\bar{\beta} \right) \Big].
\end{equation}
Here we have used the fact that $\beta_i^+ = \beta_i^+(\omega_{2i},\omega_{2i+2})$, $\beta_i^+$ is independent of $\beta_j^+$ for $j\neq i-1, i, i+1$, and same for $\beta_i^-$, due to the assumption that $\{\omega_i\}$ are independent random variables.

%Because most of $ \beta_i^{\pm} $ are independent random variables except those with close sub-index, the expectation of crossing-terms in the (\ref{varbeta}) would vanish except (\ref{sigmabeta}).

\subsection{Stochastic continuum elastic energy}

As in the previous section for the misfit energy, here we
 obtain the continuum limit of the stochastic elastic energy under the assumption that $h \ll \delta \ll L$, where $\delta$ is the size of the sumpercell and $L$ is the length scale of the continuum model.
 The elastic energies in Eq.~\eqref{atomic_elastic} depend on the summation of stochastic stiffness coefficients $ \sum_{i = 1}^n(\beta_i^{\pm} - \bar{\beta}) $.
We derive a continuum formulation of $ \sum_{i = 1}^n(\beta_i^{\pm} - \bar{\beta}) $
 by letting the size of the supercell $n\rightarrow\infty$.
    We perform numerical simulations to examine this limit in this subsection. More rigorous convergence proof using a modified central limit theorem will be given in Sec.~\ref{subsec:convergenceproof}.

\begin{figure} [tbhp]
	\centering
	\includegraphics[width=0.6\columnwidth]{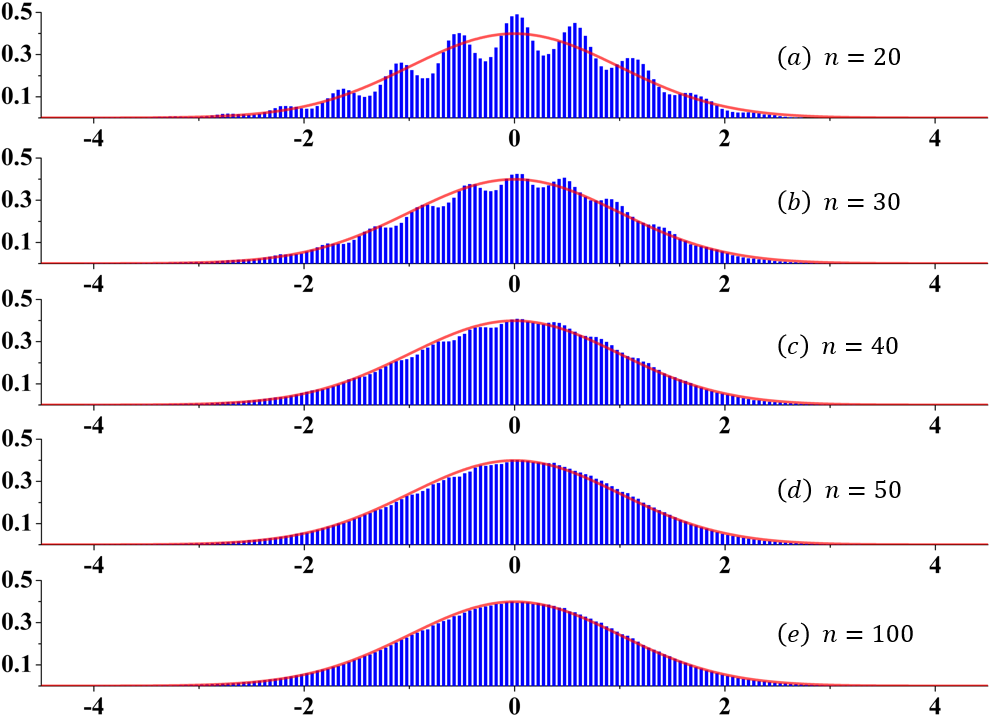}
	\caption{\label{beta_distribution} Probability distributions of the summation $ \sum_{i = 1}^n(\beta_i^{\pm} - \bar{\beta}) $ for different values of supercell size $ n $: (a) $ n = 20 $, (b) $ n = 30 $, (c) $ n = 40 $, (d) $ n = 50 $, (e) $ n = 100 $. Bar graphs are  distributions of $ 10^6 $ samples, normalized by  the variance of $ \sum_{i = 1}^n(\beta_i^{\pm} - \bar{\beta}) $  in Eq.~\eqref{varbeta}.
The red lines are the probability density function of the Gaussian distribution $\mathcal{N}(0,1)$. }
\end{figure}

In the numerical simulations, we use the same HEA system in Eq.~\eqref{example} being used  for deriving the misfit energy in the previous section, which consists of five elements  with parameters shown in Table~\ref{table_para}.  We sampled total number of $ 10^6 $ atomic configurations by the probability distribution \eqref{measureomega}  for each value of the supercell size $ n $. Each atomic configuration corresponds to a value of the summation $ \sum_{i = 1}^n(\beta_i^{\pm} - \bar{\beta}) $.
  Note that from Eq.~\eqref{beta}, the stiffness coefficients of the top or the bottom layer are functions of atom species within the layer, and hence $ \sum_{i = 1}^n(\beta_i^+ - \bar{\beta}) $ and $ \sum_{i = 1}^n(\beta_i^- - \bar{\beta}) $ are independent and identically distributed. Therefore, it is sufficient to consider the summations of either one of the top or the bottom layer.
 The normalized distributions of the sample values of $ \sum_{i = 1}^n(\beta_i^{\pm} - \bar{\beta}) $ for different values of supercell size $ n $ are shown in Fig.~\ref{beta_distribution}.

As illustrated by Fig.~\ref{beta_distribution}, when $ n $ becomes large enough, the probability distribution of the summation $ \sum_{i = 1}^n(\beta_i^{\pm} - \bar{\beta}) $ converges to that of a Gaussian distribution with mean $0$ and variance $n\sigma_{\beta\beta}$. That is,
\begin{equation} \label{statement2}
	\dfrac{1}{\sqrt{n\sigma_{\beta\beta}}} \sum_{i = 1}^n (\beta_i^{\pm} - \bar{\beta}) \longrightarrow \mathcal{N}(0,1) \ \ \ \ \ \text{as} \ \ n \longrightarrow \infty.
\end{equation}
 Rigorous convergence proof  for a general case using a modified central limit theorem will be given in Sec.~\ref{subsec:convergenceproof}.

%Although the statement (\ref{statement2}) requires $ n \longrightarrow \infty $, Figure~(\ref{beta_distribution}) illustrates that when $ n $ is in the order of 10 the summation can be well approximated by normal distributions.

We have assumed that the size of the supercell $ \delta = nh $ is much small than the length unit of the continuum model. Using the notation $Y_\delta \sim \mathcal{N}(0,\delta) $ defined in Eq.~\eqref{eqn:y_delta},
 which is the Gaussian distribution with expectation $0$ and standard deviation $\sqrt{\delta}$, and Eq.~\eqref{statement2},  the  elastic energies of the top and bottom layers in the supercell given in Eq.~\eqref{atomic_elastic} can be written as
\begin{align} \label{atomic_elastic1}
	\Delta E_{\rm elastic}^{\pm} &= nh\cdot \dfrac{1}{2}\bar{\alpha} \left( \dfrac{du^{\pm}}{dx} \right)^2 + \sqrt{nh}\cdot\dfrac{\sqrt{\sigma_{\beta\beta}/h}}{2} \left( \dfrac{du^{\pm}}{dx} \right)^2 \cdot Y_1 \nonumber \\
	&= \delta \cdot \dfrac{1}{2}\bar{\alpha} \left( \dfrac{du^{\pm}}{dx} \right)^2 + \dfrac{\sqrt{\sigma_{\beta\beta}/h}}{2} \left( \dfrac{du^{\pm}}{dx} \right)^2 \cdot Y_{\delta}.
\end{align}

As did in Sec.~\ref{sec:stochastic_misfit} for deriving the misfit energy,
the slip plane is discretized into infinite such small intervals: $ \delta_1, \delta_2, \delta_3, \cdots $, and each interval is associated with a Gaussian random variable forming the sequence $ Y_{\delta_1}, Y_{\delta_2}, Y_{\delta_3}, \cdots $. As argued in  Sec.~\ref{sec:stochastic_misfit},
$ \{Y_{\delta} \}$ are approximately mutually independent and can be regarded as independent Gaussian increments, forming the Brownian motion as given in Eq.~\eqref{brownian_motion}.
Since the size of the supercell $ \delta = nh $ is much smaller than the length unit of the continuum model, from Eq.~\eqref{atomic_elastic1}, we have the following expression for the elastic energies on the continuum level:
\begin{equation}
	dE_{\rm elastic}^{\pm} = \dfrac{1}{2}\bar{\alpha} \left( \dfrac{du^{\pm}}{dx} \right)^2 dx + \dfrac{\sqrt{\sigma_{\beta\beta}/h}}{2} \left( \dfrac{du^{\pm}}{dx} \right)^2 dB_x.
\end{equation}

This equation can be written as $dE_{\rm elastic}^{\pm} = \frac{1}{2}\bar{\alpha} \left( \dfrac{du^{\pm}}{dx} \right)^2 \bigg(dx + \varepsilon_{\rm e}\sqrt{h}\, dB_x\big)$, where the dimensionless parameter
\begin{equation}\label{eqn:epsilone}
\varepsilon_{\rm e}= \frac{\sqrt{\sigma_{\beta\beta}}}{\bar{\alpha}h}.
\end{equation}
For the bilayer HEA system (\ref{example}), it can be calculated that $\varepsilon_{\rm e}=0.0914$. %which is a small number.

\section{The Peierls-Nabarro model for HEAs} \label{section_total}
In this section, we formulate the stochastic total energy of the Peierls-Nabarro model for the bilayer HEA, and rigorously prove the convergence from the atomistic model.
 The stochastic model adopted in Ref.~\cite{theoretical1} is also  examined.

%Therefore, the variance of total energy can not be expressed as direct summation of the variance from misfit energy and elastic energy.
%In this section, we first formulate atomic-scope total energy of PN model for HEAs, then analyze the correlation between the misfit energy and the elastic energy, and finally formulate continuum-scope total energy of PN model for HEAs.

\subsection{Total energy of the supercell using atomistic model and its continuum limit}\label{sec:total_atomic}
 In the Peierls-Nabarro model for an interlayer dislocation, there will be both disregistry $\phi$ across the slip plane and elastic deformation $\{u^\pm_i\}$ within each layer. We consider the same suppercell whose  size is $nh$ as in the previous two sections (see Figs.~\ref{latticeMisfit} and \ref{latticeElastic}), and the supercell has both $\phi$ and $\{u^\pm_i\}$ (with constant $\frac{du^\pm}{dx}$ as in the previous section). Using Eqs.~\eqref{eqn:subtract}, \eqref{eqn:uvbar} and  \eqref{atomic_elastic}, \eqref{mean_beta}, \eqref{mean_alpha}, the total energy of the supercell can be calculated as
\begin{align} \label{atomic_PN_energy}
	\Delta E_{\rm PN} =& \Delta E_{\rm misfit} + \Delta E_{\rm elastic}^+ + \Delta E_{\rm elastic}^- \nonumber \\
	=& nh\cdot \left( \bar{\gamma}(\phi) + \dfrac{1}{2}\bar{\alpha}\left( \dfrac{du^+}{dx} \right)^2 + \dfrac{1}{2}\bar{\alpha}\left( \dfrac{du^-}{dx} \right)^2 \right) \nonumber \\
	+& \sum_{i = 1}^{n}\Big[ (U_i - \bar{U}) + (V_i - \bar{V}) + (P_i - \bar{P}) + (Q_i - \bar{Q}) \Big],
\end{align}
in which the first term is the average value of the total energy and the second term is a stochastic contribution whose mean value is $0$. Here $\bar{P}=\frac{1}{2} \bar{\beta} \left( \frac{d u^+(x)}{d x} \right)^2$ and $\bar{Q}=\frac{1}{2} \bar{\beta} \left( \frac{d u^-(x)}{d x} \right)^2$  from Eqs.~\eqref{Pi},  \eqref{Qi}, and  \eqref{mean_beta}.

The variance of this total energy is
\begin{flalign} \label{var_total}
	&\mathbb{E} \Big( \sum_{i = 1}^{n}\big[ (U_i - \bar{U}) + (V_i - \bar{V}) + (P_i - \bar{P}) + (Q_i - \bar{Q}) \big] \Big)^2   \\
	= &\mathbb{E} \Big[\sum_{i = 1}^n (U_i - \bar{U})\Big]^2 + \mathbb{E} \Big[\sum_{i = 1}^n (V_i - \bar{V})\Big]^2 + \mathbb{E} \Big[\sum_{i = 1}^n (P_i - \bar{P})\Big]^2 + \mathbb{E} \Big[\sum_{i = 1}^n (Q_i - \bar{Q})\Big]^2   \nonumber  \\
	&+ 2\sum_{i,j=1}^{n} \Bigg\{\mathbb{E} \Big[ (U_i - \bar{U})(V_j - \bar{V}) \Big] + \mathbb{E} \Big[ (P_i - \bar{P})(Q_j - \bar{Q}) \Big]
	+ \mathbb{E} \Big[ (U_i - \bar{U})(P_j - \bar{P}) \Big]\nonumber\\
&+\mathbb{E} \Big[ (U_i - \bar{U})(Q_j - \bar{Q}) \Big]
	+  \mathbb{E} \Big[ (V_i - \bar{V})(P_j - \bar{P}) \Big] +  \mathbb{E} \Big[ (V_i - \bar{V})(Q_j - \bar{Q}) \Big]\Bigg\}. \nonumber
\end{flalign}

In Sec.~\ref{subsec:gsf}, we have calculated the variances of those terms of the misfit energy (Eq.~\eqref{var_msf}):
\begin{flalign}  \label{var_misfit}
	 &\mathbb{E} \Big[\sum_{i = 1}^n (U_i - \bar{U})\Big]^2 + \mathbb{E} \Big[\sum_{i = 1}^n (V_i - \bar{V})\Big]^2 + 2\sum_{i,j=1}^{n} \mathbb{E} \Big[ (U_i - \bar{U})(V_j - \bar{V}) \Big]\\
=&\mathbb{E} \Big[\sum_{i = 1}^n \big((U_i - \bar{U})+(V_i - \bar{V})\big)\Big]^2
 = nh \cdot \theta^2(\phi).\nonumber
\end{flalign}
Using the variances of the elastic energies in the top and bottom layers calculated in Sec.~\ref{subsec:elastic-mean-var} (Eqs.~\eqref{eqn:elastic0}, \eqref{atomic_elastic}, and \eqref{varbeta}), we have
\begin{subequations}
	\begin{align}
		\mathbb{E} \Big[\sum_{i = 1}^n (P_i - \bar{P})\Big]^2 = \dfrac{1}{4} \left( \dfrac{du^+}{dx} \right)^4 \mathbb{E} \Big[\sum_{i = 1}^n (\beta_i^+ - \bar{\beta})\Big]^2 = \dfrac{1}{4} \left( \dfrac{du^+}{dx} \right)^4 \cdot n\sigma_{\beta\beta},\\
		\mathbb{E} \Big[\sum_{i = 1}^n (Q_i - \bar{Q})\Big]^2 = \dfrac{1}{4} \left( \dfrac{du^-}{dx} \right)^4 \mathbb{E} \Big[\sum_{i = 1}^n (\beta_i^- - \bar{\beta})\Big]^2 = \dfrac{1}{4} \left( \dfrac{du^-}{dx} \right)^4 \cdot n\sigma_{\beta\beta}.
	\end{align}
\end{subequations}
Since the atomic configurations of the top and the bottom layers are mutually independent,  the covariance of their elastic energies  vanishes:
\begin{equation}
	\mathbb{E} \Big[ (P_i - \bar{P})(Q_j - \bar{Q}) \Big] = \mathbb{E} (P_i - \bar{P}) \cdot \mathbb{E} (Q_j - \bar{Q}) = 0.
\end{equation}

The remaining part in Eq.~\eqref{var_total} (sum of the last four terms) is the covariance between the misfit energy and the elastic energy.
The covariances between different terms of the misfit energy and the elastic energy can be calculated as
{\small	
\begin{subequations} \label{var_msf-elas}
	\begin{align}
	\sum_{i,j=1}^{n} \mathbb{E} \Big[ (U_i - \bar{U})(P_j - \bar{P}) \Big] = \frac{1}{2}\left( \frac{du^+}{dx} \right)^2  n \sigma_{\beta u}(\phi), \ \
		\sum_{i,j=1}^{n} \mathbb{E} \Big[ (U_i - \bar{U})(Q_j - \bar{Q}) \Big] = \frac{1}{2}\left( \frac{du^-}{dx} \right)^2  n \sigma_{\beta u}(\phi), \\
		\sum_{i,j=1}^{n} \mathbb{E} \Big[ (V_i - \bar{V})(P_j - \bar{P}) \Big] = \frac{1}{2}\left( \frac{du^+}{dx} \right)^2  n \sigma_{\beta v}(\phi), \ \
		\sum_{i,j=1}^{n} \mathbb{E} \Big[ (V_i - \bar{V})(Q_j - \bar{Q}) \Big] = \frac{1}{2}\left( \frac{du^-}{dx} \right)^2  n \sigma_{\beta v}(\phi).
	\end{align}
\end{subequations}
}
where
\begin{equation} \label{sigma_betauv}
			\sigma_{\beta u}(\phi) := \mathbb{E} \Big[ (U_i - \bar{U})(\beta_{i-1}^{\pm} + \beta_i^{\pm} - 2 \bar{\beta}) \Big], \ \
		\sigma_{\beta v}(\phi) := \mathbb{E} \Big[ (V_i - \bar{V})(\beta_{i-1}^{\pm} + \beta_i^{\pm} - 2 \bar{\beta}) \Big].
\end{equation}
Here, similar to the calculation of $\sigma_{uv}(\phi)$ in Eq.~\eqref{sigma_uv}, we have used the property that $U_i$ is not independent only of $P_{i-1}$ and $P_{i}$ (i.e., $\beta^+_{i-1}$ and $\beta^+_{i}$) and same for other covariances.

Summarizing Eqs.~\eqref{var_misfit}--\eqref{var_msf-elas}, the variance of this total energy of the supercell in Eq.~\eqref{var_total} can be written as
\begin{equation}
	 \mathbb{E} \Big( \sum_{i = 1}^{n}\big[ (U_i - \bar{U}) + (V_i - \bar{V}) + (P_i - \bar{P}) + (Q_i - \bar{Q}) \big] \Big)^2=nh \cdot \sigma^2 \Big( \phi,\dfrac{du^+}{dx}, \dfrac{du^-}{dx} \Big),
\end{equation}
where
\begin{equation}\label{eqn:var_total}
	\sigma^2 \Big( \phi,\dfrac{du^+}{dx}, \dfrac{du^-}{dx} \Big) :=
		 \theta^2(\phi) + \dfrac{\sigma_{\beta\beta}}{4h} \left[ \left(\dfrac{du^+}{dx}\right)^4 + \left( \dfrac{du^-}{dx} \right)^4 \right] +\eta(\phi) \left[ \left(\dfrac{du^+}{dx}\right)^2 + \left( \dfrac{du^-}{dx} \right)^2 \right],
\end{equation}
\begin{equation} \label{eta}
	\eta(\phi) := \dfrac{1}{h} \Big( \sigma_{\beta u}(\phi) + \sigma_{\beta v}(\phi) \Big).
\end{equation}

%We further define a notation to combine $ \sigma_{\beta u} $ and $ \sigma_{\beta v} $ as

%There exists a co-factor $ n $ in all the variances and covariances evaluated from (\ref{var_misfit}) to (\ref{var_msf-elas}). Therefore, we introduce the averaged standard derivation $ \sigma \Big( \phi,\dfrac{du^+}{dx}, \dfrac{du^-}{dx} \Big) $ to express the total variance as
%Since the variances and covariances evaluated in (\ref{var_misfit})-(\ref{var_msf-elas}) all co-exist a factor $ n $, we introduce the averaged standard derivation $ \sigma \Big( \phi,\dfrac{du^+}{dx}, \dfrac{du^-}{dx} \Big) $, such that the total variance can be expressed as
%
%With all the variances and covariances in (\ref{var_total}) being expressed from (\ref{var_misfit}) to (\ref{var_msf-elas}), the averaged standard derivation $ \sigma \Big( \phi,\dfrac{du^+}{dx}, \dfrac{du^-}{dx} \Big) $ can be evaluated as
%

Similar to the continuum limits of the misfit energy in Eq.~\eqref{statement1} (shown in Fig.~\ref{gammadistribution}) and the elastic energy in Eq.~\eqref{statement2} (shown in Fig.~\ref{beta_distribution}), numerical simulations also suggest that the stochastic perturbation in the total energy $\Delta E_{\rm PN}$ in Eq.~\eqref{atomic_PN_energy}
%As is suggested by the calculation example that the normalized distributions of local misfit energy (Figure~\ref{gammadistribution}) and local elastic energy (Figure~\ref{beta_distribution}) both converge to standard normal distributions, the inflation of total energy are also expected to
converges to a Gaussian distribution:
\begin{equation} \label{statement3}
	\dfrac{\sum\limits_{i = 1}^{n}\Big[ (U_i - \bar{U}) + (V_i - \bar{V}) + (P_i - \bar{P}) + (Q_i - \bar{Q}) \Big]}{\sqrt{nh}\cdot\sigma\left(\phi,\dfrac{du^+}{dx},\dfrac{du^-}{dx}\right)} \longrightarrow \mathcal{N}(0,1), \ {\rm as} \ n \longrightarrow \infty.
\end{equation}
This limit will be proved in the next subsection.
 When $ du^{\pm}/dx = 0 $, this limit reduces to the continuum limit of the misfit energy in Eq.~\eqref{statement1}. When $ \phi = 0 $ and only the elastic energy of either the top or the bottom layer is considered, this limit reduces to Eq.~\eqref{statement2}.

\begin{figure} [tbhp]
	\centering
\subfigure[]{	\includegraphics[width=0.45\columnwidth]{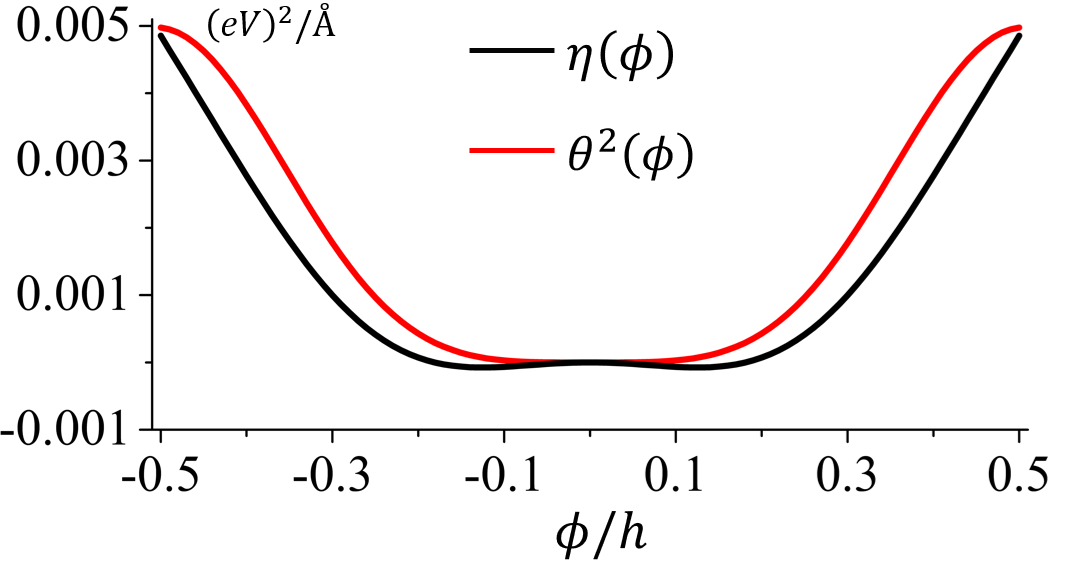}}
\subfigure[]{   \includegraphics[width=0.45\columnwidth]{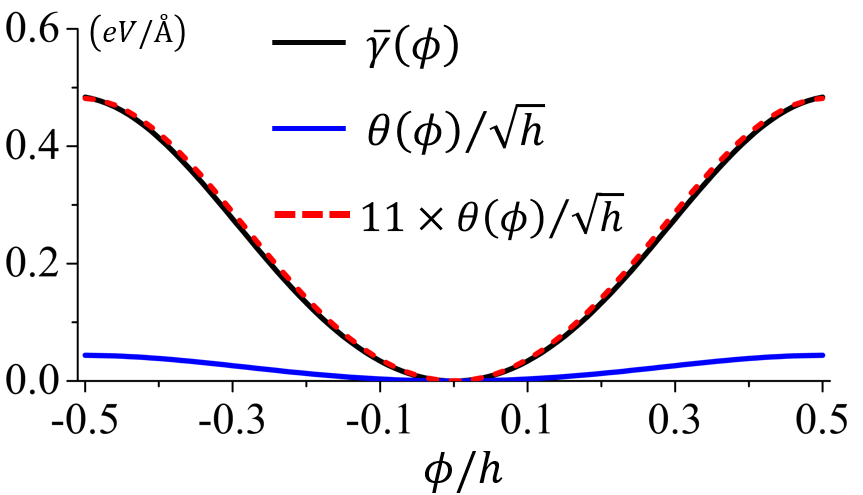} }
	\caption{\label{thetaeta} (a) Functions $ \theta^2(\phi) $ and $ \eta(\phi) $ in the standard deviation of the total energy density $\sigma \Big( \phi,\frac{du^+}{dx}, \frac{du^-}{dx} \Big)  $ defined in Eq.~\eqref{eqn:var_total}, calculated using the  bilayer HEA system  (\ref{example}). (b) Comparison of $\theta(\phi)$ with the $\gamma$-surface $\bar{\gamma}(\phi)$.}
\end{figure}

 The standard deviation of the total energy density $\sigma \Big( \phi,\frac{du^+}{dx}, \frac{du^-}{dx} \Big)  $ defined in Eq.~\eqref{eqn:var_total} depends on the elastic strain $ du^{\pm}/dx $ in the top and the bottom layers  and on the disregistry $\phi$ between the two layers  through functions $ \theta(\phi) $ and $ \eta(\phi) $, where  $ \theta(\phi) $ is the standard deviation of the misfit energy (see Eq.~\eqref{statement1-1}) and  $ \eta(\phi) $ defined in Eqs.~\eqref{var_msf-elas} and \eqref{eta}  is associated with the covariance between the elastic energy and misfit energy.

  For the  bilayer HEA system  (\ref{example}), the calculated functions $ \theta^2(\phi) $ and $ \eta(\phi) $ are shown in Fig.~\ref{thetaeta}(a).
We also compare the function $\theta(\phi)$ with the gamma surface $\bar{\gamma}(\phi)$ using the  bilayer HEA system  (\ref{example}), see Fig.~\ref{thetaeta}(b). It can be seen that we can have the following relation
\begin{equation}\label{theta:perturbation}
\theta(\phi)=\varepsilon_{\rm m} \sqrt{h}\bar{\gamma}(\phi),
\end{equation}
for some small $\varepsilon_{\rm m}$. Here we can choose $\varepsilon=1/11$.
%Using Eqs.~\eqref{eqn:epsilone} and \eqref{theta:perturbation}, the variance in Eq.~\eqref{eqn:var_total} can be written as
%\begin{equation}\label{eqn:var_total1}
%	\sigma^2 \Big( \phi,\dfrac{du^+}{dx}, \dfrac{du^-}{dx} \Big) :=
%		 \theta^2(\phi) + \dfrac{\sigma_{\beta\beta}}{4h} \left[ \left(\dfrac{du^+}{dx}\right)^4 + \left( \dfrac{du^-}{dx} \right)^4 \right] +\eta(\phi) \left[ \left(\dfrac{du^+}{dx}\right)^2 + \left( \dfrac{du^-}{dx} \right)^2 \right],
%\end{equation}

\subsection{Prove of convergence to Gaussian distribution}\label{subsec:convergenceproof}
In probability theory, the central limit theorem states that the normalized summation of independent random variables tends towards a normal distribution as the number of random variables goes to infinity. However, the random variables $\{ (U_i - \bar{U}) + (V_i - \bar{V}) + (P_i - \bar{P}) + (Q_i - \bar{Q}) \} $ in the summation in Eq.~\eqref{statement3} are not mutually independent when the sub-index $ i $ varies. Thus the central limit theorem does not apply to it directly.  A modified central limit theorem still holds when the assumption of independence in classical central limit theorem is relaxed to weak dependence \cite{probability1}. In this subsection, we  apply the modified central limit theorem to prove the convergence in Eq.~\eqref{statement3} (and accordingly the convergence in Eqs.~\eqref{statement1} and \eqref{statement2} as two special cases).

The weak dependence means that the random variables in a sequence far apart from one another are nearly independent \cite{probability2}, which is called $ \alpha $-mixing and is measured by a mixing coefficient. For the random variable sequence $ \{X_i\}_{i=1}^{\infty} $, the mixing coefficient $ \alpha_n $ is defined as
\begin{equation}
	\alpha_n = \sup \Big\{ \left| P(A \cap B) - P(A)P(B) \right|: \forall k = 1,2,\cdots,\ A\in \mathcal{F}_1^k,\ B\in\mathcal{F}_{k+n}^{\infty}   \Big\}  ,
\end{equation}
in which $ \mathcal{F}_a^b $ denotes the $ \sigma $-field generated by $ \{ X_a,X_{a+1},\cdots,X_b \} $. Suppose that $ \alpha_n \rightarrow 0 $, then $ X_k $ and $ X_{k+n} $ are approximately independent for large $ n $ uniformly over all $ k $. With the definition of mixing coefficient, the modified central limit theorem holds for a weakly-dependent random-variable sequence \cite{probability1}.
\begin{theorem}\cite{probability1}	 \label{clt}
Suppose that random variables $ X_1, X_2, \cdots  $ are stationary with $ \alpha $-mixing coefficient $ \alpha_n = O\left( n^{-5} \right) $, and  $ \mathbb{E} (X_n) = 0 $, $ \mathbb{E} \left[ X_n^{12} \right] < \infty $. Let $ S_n = X_1 + \cdots + X_n $ and  $ \sigma^2 = \lim\limits_{n\rightarrow\infty} \mathbb{E} \left[ S_n^2 \right] / n $, where $ \sigma $ is positive, then
	\begin{equation} \label{thm}
		\dfrac{S_n}{\sigma \sqrt{n}} \longrightarrow \mathcal{N}(0,1), \ \ {\rm as}\ n\rightarrow \infty.
	\end{equation}
\end{theorem}

To prove the convergence  in Eq.~\eqref{statement3}, we set $ X_i = (U_i - \bar{U}) + (V_i - \bar{V}) + (P_i - \bar{P}) + (Q_i - \bar{Q}) $. Obviously, the sequence $ \{ X_i \}_{i=1}^{\infty} $ is stationary, $ \mathbb{E} (X_i) = 0 $ and $ \mathbb{E} \left[ X_i^{12} \right] < \infty $. We now check the $ \alpha $-mixing coefficient.

Note that the energy components $ U_i $, $ V_i $, $ P_i $ and $ Q_i $ are defined based on the local atomic configurations $ \omega_{2i-1} $, $ \omega_{2i} $, $ \omega_{2i+1} $ and $ \omega_{2i+2} $. Thus  $ X_i $ is  independent with $ X_{i\pm n} $ when $ n\geq 2 $. Therefore, in our case, for $ k = 1,2,\cdots $, and $ \forall A\in\mathcal{F}_1^k, \forall B\in\mathcal{F}_{k+n}^{\infty} $, the mixing coefficient
\begin{equation}
	\alpha_n = \sup\left|P(A\cap B) - P(A)P(B)\right| = 0 \leq O(n^{-5}) \ \ \ \text{when}\ n\geq 2.
\end{equation}
The condition of the modified central limit theorem holds. The convergence in Eq.~\eqref{statement3} follows from the
 conclusion of the theorem in Eq.~\eqref{thm}. The convergence in Eqs.~\eqref{statement1} and \eqref{statement2} hold accordingly as special cases.

\subsection{Stochastic total energy}
%Although the statement (\ref{statement3}) require $ n \longrightarrow \infty $, it is sufficient to approximate the inflation term as normal random variable for finite $ n $. By setting $ \delta = nh $ and defining $ Y_\delta \sim \mathcal{N}(0,1) $, the atomic-scope total energy (\ref{atomic_PN_energy}) of the interval $ nh $ can be re-expressed as
From Eq.~\eqref{statement3}, as  $ n \longrightarrow \infty $,  the  total energy (\ref{atomic_PN_energy}) of the supercell with $ \delta=nh $ can be written as
\begin{align} \label{atomic_PN_energy1}
	\Delta E_{\rm PN} =& nh\cdot \left( \bar{\gamma}(\phi) + \dfrac{1}{2}\bar{\alpha}\left( \dfrac{du^+}{dx} \right)^2 + \dfrac{1}{2}\bar{\alpha}\left( \dfrac{du^-}{dx} \right)^2 \right) + \sqrt{nh}\cdot \sigma\left( \phi,\dfrac{du^+}{dx},\dfrac{du^-}{dx} \right)\cdot Y_1 \nonumber\\
	=& \delta \cdot \left( \bar{\gamma}(\phi) + \dfrac{1}{2}\bar{\alpha}\left( \dfrac{du^+}{dx} \right)^2 + \dfrac{1}{2}\bar{\alpha}\left( \dfrac{du^-}{dx} \right)^2 \right) + \sigma\left( \phi,\dfrac{du^+}{dx},\dfrac{du^-}{dx} \right)\cdot Y_{\delta}.
\end{align}
Recall that $ Y_\delta \sim \mathcal{N}(0,1) $.
As in the continuum limit in previous sections,
the slip plane is divided into infinite such small intervals: $ \delta_1,\delta_2,\delta_3,\cdots $, and each interval is associated with a Gaussian random variable forming a sequence $ Y_{\delta1}, Y_{\delta_2}, Y_{\delta_3},\cdots $, which are  independent Gaussian increments and form the Brownian motion as described in Eq.~\eqref{brownian_motion}. Using the assumption that
 $ \delta $ is small compared with the length unit in the continuum model,  the continuum limit of Eq.~\eqref{atomic_PN_energy1}, using integral form, is:
\begin{equation} \label{total_energy}
	E_{\rm PN} =\int_{-\infty}^{+\infty} \left( \bar{\gamma}(\phi) + \dfrac{1}{2}\bar{\alpha}\left( \dfrac{du^+}{dx} \right)^2 + \dfrac{1}{2}\bar{\alpha}\left( \dfrac{du^-}{dx} \right)^2 \right) dx + \sigma\left( \phi,\dfrac{du^+}{dx},\dfrac{du^-}{dx} \right) dB_x.
\end{equation}

%or in integral form:
%\begin{align} \label{total_energy}
%	E_{PN}[\phi,u^+,u^-] = \int_{-\infty}^{+\infty}&\left( \bar{\gamma}(\phi) + \dfrac{1}{2}\bar{\alpha}\left| \dfrac{du^+}{dx} \right|^2 + \dfrac{1}{2}\bar{\alpha}\left| \dfrac{du^-}{dx} \right|^2 \right) dx   \nonumber\\
%	+& \int_{-\infty}^{+\infty} \sigma\left( \phi,\dfrac{du^+}{dx},\dfrac{du^-}{dx} \right) dB_x.
%\end{align}

Recall that in this formula, $ \phi(x) $ is the diregistry across the slip plane, and $ u^+(x) $, $ u^-(x) $ are displacements in the upper and lower layers, respectively.
They have the relation
%\begin{equation}
	$\phi(x) = u^+(x) - u^-(x)$.
%\end{equation}
In the Peierls-Nabarro models \cite{PN1,PN2}, it is assumed that $u^+(x) = -u^-(x)$, and accordingly,
$u^+(x) =- u^-(x)=\frac{1}{2}\phi(x)$ from the  equation above. Under these conditions, the total energy in  Eq.~\eqref{total_energy} can be written as an expression that depends only on $\phi(x)$:
\begin{equation} \label{total_energy1}
	E_{\rm PN} = \int_{-\infty}^{+\infty}\left( \bar{\gamma}(\phi) + \dfrac{1}{4}\bar{\alpha}\left( \dfrac{d\phi}{dx} \right)^2\right) dx + \bar{\sigma}\left( \phi,\dfrac{d\phi}{dx} \right) dB_x,
\end{equation}
where from Eq.~\eqref{eqn:var_total},
%\begin{equation} \label{total_energy2}
	$\bar{\sigma}\left( \phi,\frac{d\phi}{dx} \right)=\sqrt{
		 \theta^2(\phi) + \frac{1}{32h} \sigma_{\beta\beta}\left(\frac{d\phi}{dx}\right)^4  +\frac{1}{2}\eta(\phi)  \left(\frac{d\phi}{dx}\right)^2}$.
%\end{equation}

If we consider the randomness in the misfit energy and elastic energies separately as in previous two sections, we have the following formulation for the total energy of the Peierls-Nabarro model:
\begin{flalign}\label{eqn:energynew1}
E_{\rm PN} =& \int_{-\infty}^{+\infty} \bar{\gamma}(\phi)\big(dx + \varepsilon_{\rm m}\sqrt{h} \, dB^{(1)}_x\big) \nonumber\\
	& +\dfrac{1}{2}\bar{\alpha}\left( \dfrac{du^+}{dx} \right)^2 \left(dx + \varepsilon_{\rm e}\sqrt{h} \, dB^{(2+)}_x\right)+\dfrac{1}{2}\bar{\alpha}\left( \dfrac{du^-}{dx} \right)^2 \left(dx + \varepsilon_{\rm e}\sqrt{h} \, dB^{(2-)}_x\right),
\end{flalign}
where the Brownian motion $ B^{(1)}_x$, $B^{(2+)}_x $ and $ B^{(2-)}_x $ represent the randomness in the misfit energy and the elastic energies of the top and bottom layers, respectively. Because the randomness in each energy component correspond to the same random atomic configuration, these Brownian motions are not mutually independent. Using the covariances of different energies on the atomic level calculated in Sec.~\ref{sec:total_atomic}, we obtain the covariances between these Brownian motions  as follows. For any $ s_1 \leq s_2, \tau_1 \leq \tau_2 $, using the notation $ \delta_c $ as the length of the overlap between the two open sets $ (s_1,s_2) $ and $ (\tau_1,\tau_2) $, the correlations are
\begin{flalign}
	&{\rm Cov}(B_{s_2}^{(1)} - B_{s_1}^{(1)},B_{\tau_2}^{(2\pm)} - B_{\tau_1}^{(2\pm)}) = \sigma_{\rm em}\delta_c \\
%	Cov(B_{s_2}^{(1)} - B_{s_1}^{(1)},B_{\tau_2}^{(3)} - B_{\tau_1}^{(3)}) =& \dfrac{\eta(\phi)}{\theta(\phi)\cdot\sqrt{\sigma_{\beta\beta}/h}}\cdot \varepsilon \\
	&{\rm Cov}(B_{s_2}^{(2+)} - B_{s_1}^{(2+)},B_{\tau_2}^{(2-)} - B_{\tau_1}^{(2-)}) = 0,
\end{flalign}
where
\begin{equation}\label{eqn:sigma_em}
 \sigma_{\rm em}=\frac{\eta(\phi)}{\theta(\phi)\cdot\sqrt{\sigma_{\beta\beta}/h}}
=\frac{\eta(\phi)}{\varepsilon_{\rm e}\varepsilon_{\rm m}h\bar{\alpha}\bar{\gamma}(\phi)}.
  \end{equation}
Here we have used the small parameters $\varepsilon_{\rm e}$ and $\varepsilon_{\rm m}$ defined in Eqs.~\eqref{eqn:epsilone} and \eqref{theta:perturbation}. This energy formulation is an alternative form of Eq.~\eqref{total_energy}.

When $ u^+ = - u^-$ in the Peierls-Nabarro model, the total energy is
\begin{equation}\label{eqn:energynew2}
E_{\rm PN} = \int_{-\infty}^{+\infty} \bar{\gamma}(\phi)\big(dx + \varepsilon_{\rm m}\sqrt{h} \, dB^{(1)}_x\big)
	 +\dfrac{1}{4}\bar{\alpha}\left( \dfrac{d\phi}{dx} \right)^2 \left(dx + \varepsilon_{\rm e}\sqrt{h} \, dB^{(2)}_x\right),
\end{equation}
where the Brownian motion $ B^{(1)}_x $ and $ B^{(2)}_x $ represent the randomness in the misfit energy and the elastic energy, respectively, and the covariance between them is
\begin{equation}
	{\rm Cov}(B_{s_2}^{(1)} - B_{s_1}^{(1)},B_{\tau_2}^{(2)} - B_{\tau_1}^{(2)}) =\sqrt{2} \sigma_{\rm em}\delta_c, \end{equation}
where the notations $ s_1, s_2, \tau_1, \tau_2 $ and $ \delta_c $ are  the same as specified above.
This energy formulation is an alternative form of Eq.~\eqref{total_energy1}.

\subsection{Smoothed  stochastic total energy}
Using the stochastic energy in Eq.~\eqref{eqn:energynew1} or \eqref{eqn:energynew2} (or the formulation in Eq.~\eqref{total_energy} or \eqref{total_energy1} using a single Brownian motion), we have a Dirac delta function-like energy density and accordingly infinite point force in the Peierls-Nabarro model, which is not practical to describe the continuum profile of the dislocation core structure. On the other hand, resolution in the continuum Peierls-Nabarro model below atomic distance is not physically meaningful. Based on these, we make average over the size of an atomic site in the obtained continuum models as follows.

We first consider the misfit energy:
\begin{flalign}
	E_{\rm misfit} =& \int_{-\infty}^{+\infty} \bar{\gamma}(\phi(x))dx + \int_{-\infty}^{+\infty}\bar{\gamma}(\phi(x))\varepsilon_{\rm m}\sqrt{h} \, d B_x^{(1)} \nonumber\\
	=& \int_{-\infty}^{+\infty} \bar{\gamma}(\phi(x))dx + \sum_{n}\int_{na}^{(n+1)a} \bar{\gamma}(\phi(x))\varepsilon_{\rm m}\sqrt{h} \, d B_x^{(1)} \nonumber\\
	\approx& \int_{-\infty}^{+\infty} \bar{\gamma}(\phi(x))dx + \sum_{n} \bar{\gamma}(\phi(na))\varepsilon_{\rm m}\sqrt{h} \, \int_{na}^{(n+1)a} d B_x^{(1)} \nonumber\\
	=& \int_{-\infty}^{+\infty} \bar{\gamma}(\phi(x))dx + \sum_{n} \bar{\gamma}(\phi(na))\varepsilon_{\rm m} \frac{B_{na + a}^{(1)} - B_{na}^{(1)}}{\sqrt{h}}\cdot h \nonumber\\
\approx &\int_{-\infty}^{+\infty} \bar{\gamma}(\phi(x))dx + \int_{-\infty}^{+\infty}\bar{\gamma}(\phi(x))\varepsilon_{\rm m} \cdot Y_1^{(1)}(x,\omega)dx,\nonumber\\
= &\int_{-\infty}^{+\infty} \bar{\gamma}(\phi(x))\bigg(1+\varepsilon_{\rm m} Y_1^{(1)}(x,\omega)\bigg)dx.
\end{flalign}
Here the stochastic process $ Y_1^{(1)}(x,\omega) $  describes the increment of Brownian motion:
%\begin{equation} \label{deff1}
$	Y_1^{(1)}(x,\omega) = \frac{B_{x + h}^{(1)} - B_{x}^{(1)}}{\sqrt{h}}$,
%\end{equation}
which has the properties $ Y_1^{(1)}(x,\omega) \sim \mathcal{N}(0,1) $, and  $ Y_1^{(1)}(x_1,\omega)$,  $ Y_1^{(1)}(x_2,\omega)$ are independent when $x_1\neq x_2$.

Performing similar average in the elastic energy $E_{\rm elastic}$, we have the smoothed stochastic total energy
\begin{flalign}
E_{\rm PN} =& \int_{-\infty}^{+\infty} \bar{\gamma}(\phi)\bigg(1+\varepsilon_{\rm m} Y_1^{(1)}(x,\omega)\bigg)dx  +\dfrac{1}{2}\bar{\alpha}\left( \dfrac{du^+}{dx} \right)^2 \bigg(1+\varepsilon_{\rm e} Y_1^{(2+)}(x,\omega)\bigg)dx\nonumber \\
	&
+\dfrac{1}{2}\bar{\alpha}\left( \dfrac{du^-}{dx} \right)^2 \bigg(1+\varepsilon_{\rm e} Y_1^{(2-)}(x,\omega)\bigg)dx.
\end{flalign}
Here
%\begin{equation} \label{deff12}
$Y_1^{(2\pm)}(x,\omega) = \frac{B_{x + h}^{(2\pm)} - B_{x}^{(2\pm)}}{\sqrt{h}}$,
%\end{equation}
which has the properties $ Y_1^{(2\pm)}(x,\omega) \sim \mathcal{N}(0,1) $, and  $ Y_1^{(2\pm)}(x_1,\omega)$,  $ Y_1^{(2\pm)}(x_2,\omega)$ are independent when $x_1\neq x_2$.
The covariances of $ Y_1^{(1)}(x,\omega)$,  $Y_1^{(2+)}(x,\omega)$, and $Y_1^{(2-)}(x,\omega)$ are
%\begin{flalign}
	${\rm Cov}\big(Y_1^{(1)}(x,\omega),Y_1^{(2\pm)}(x,\omega)\big) = \sigma_{\rm em}$ and
${\rm Cov}\big(Y_1^{(2+)}(x,\omega),Y_1^{(2-)}(x,\omega)\big) = 0$,
%\end{flalign}
where $\sigma_{\rm em}$ is defined in Eq.~\eqref{eqn:sigma_em}.
Note that since $ Y_1^{(2\pm)}(x,\omega) \sim \mathcal{N}(0,1) $, and  $ Y_1^{(2\pm)}(x_1,\omega)$,  $ Y_1^{(2\pm)}(x_2,\omega)$ all have Gaussian distribution $\mathcal{N}(0,1) $, their correlations are
%\begin{equation}
$\rho\big(Y_1^{(1)}(x,\omega),Y_1^{(2\pm)}(x,\omega)\big) = \sigma_{\rm em}$ and
	$\rho\big(Y_1^{(2+)}(x,\omega),Y_1^{(2-)}(x,\omega)\big) = 0$.
%\end{equation}

When $ u^+ = - u^-$ in the Peierls-Nabarro model, this total energy becomes
\begin{equation}\label{eqn:final}
E_{\rm PN} = \int_{-\infty}^{+\infty} \bar{\gamma}(\phi)\bigg(1+\varepsilon_{\rm m} Y_1^{(1)}(x,\omega)\bigg)dx
	 +\dfrac{1}{4}\bar{\alpha}\left( \dfrac{d\phi}{dx} \right)^2\bigg(1+\varepsilon_{\rm e} Y_1^{(2)}(x,\omega)\bigg)dx,
\end{equation}
where the random variables $ Y_1^{(1)}(x,\omega), Y_1^{(2)}(x,\omega)\sim \mathcal{N}(0,1) $ represent the randomness in the misfit energy and the elastic energy, respectively.  These Gaussian random variables are independent at different locations, and the correlation and covariance between them are
%\begin{equation}
	$\rho\big(Y_1^{(1)}(x,\omega),Y_1^{(2)}(x,\omega)\big)={\rm Cov}\big(Y_1^{(1)}(x,\omega),Y_1^{(2)}(x,\omega)\big) = \sigma_{\rm em}$.
%\end{equation}

 In Ref.~\cite{theoretical1}, the stochastic effects in the nonlinear interaction associated with the dislocation core under the Peierls-Nabarro model are incorporated phenomenologically by a stochastic misfit energy, which is in the form of $E_{\rm misfit}=\int^{+\infty}_{-\infty} \eta(x)\bar{\gamma}(\phi)dx$ with $\eta(x)$ being a random variable at each location $x$ (Eq.~(8) of \cite{theoretical1}, with slightly different notations). In the stochastic Peierls-Nabarro model in Eq.~\eqref{eqn:final} obtained here, if we only consider the misfit energy, it is $E_{\rm misfit}=\int_{-\infty}^{+\infty} \big(1+\varepsilon_{\rm m} Y_1^{(1)}(x,\omega)\big)\bar{\gamma}(\phi)dx$.
Perfect agreement can be seen if we choose $\eta(x)=1+\varepsilon_{\rm m} Y_1^{(1)}(x,\omega)$ in the stochastic model in  Ref.~\cite{theoretical1}. This validates the stochastic model adopted in  Ref.~\cite{theoretical1}.

\section{Summary} \label{section_summary}
%%%%%%%%%%%%%%%%%%%%%%%%%%%%%%%%%%%%%%%%%%%%%%%
%In this paper, we  derive a stochastic continuum model based on the Peierls-Nabarro framework for inter-layer dislocations in a bilayer
%HEA from an atomistic model that incorporates the atomic level randomness.  The obtained continuum model can be considered as a stochastic generalization of the classical, deterministic Peierls-Nabarro model for the dislocation core and related properties.
%This derivation also validates the stochastic model adopted by Zhang $et$ $al.$ (Acta Mater. 166, 424-434, 2019).

We have derived a continuum model for inter-layer dislocations in a bilayer HEA from an atomistic model that incorporates the atomic level randomness. The continuum model is under the framework of the Peierls-Nabarro model, in which the nonlinear effect within the dislocation core region is included.
The obtained continuum stochastic total energy can be written in the form of either a single Brownian motion or multiple Brownian motions (separating the stochastic effects in different energies).
Smoothed formulations of the stochastic total energy are also presented. The derivation validates the stochastic model adopted in Ref.~\cite{theoretical1}.

%\section*{Acknowledgments}
%We would like to acknowledge xxx.

\bibliographystyle{siamplain}
\bibliography{references}
\end{document}